\documentclass[runningheads]{llncs}

\usepackage{amsmath,amssymb,amsfonts, mathtools}
\usepackage{algorithm, algpseudocode}
\usepackage{graphicx}
\usepackage{textcomp}
\usepackage{xcolor}
\usepackage{booktabs} 
\usepackage{hyperref}
\usepackage{amsmath}
\usepackage{enumitem}
\usepackage{relsize}
\usepackage{subcaption}
\usepackage{multirow}
\usepackage[depth=3]{bookmark}

\DeclarePairedDelimiter{\ceil}{\lceil}{\rceil}

\begin{document}
\title{Experimental Analysis of Locality Sensitive Hashing Techniques for High-Dimensional Approximate Nearest Neighbor Searches}
%
%
\author{Omid Jafari\orcidID{0000-0003-3422-2755} \and
Parth Nagarkar\orcidID{0000-0001-6284-9251}}
\authorrunning{O. Jafari et al.}
%
\institute{New Mexico State University, Las Cruces, US \and
\email{\{ojafari, nagarkar\}@nmsu.edu}}
\maketitle              
\begin{abstract}
Finding nearest neighbors in high-dimensional spaces is a fundamental operation in many multimedia retrieval applications. Exact tree-based indexing approaches are known to suffer from the notorious \textit{curse of dimensionality} for high-dimensional data. Approximate searching techniques sacrifice some accuracy while returning \textit{good enough} results for faster performance. Locality Sensitive Hashing (LSH) is a very popular technique for finding approximate nearest neighbors in high-dimensional spaces. Apart from providing theoretical guarantees on the query results, one of the main benefits of LSH techniques is their good scalability to large datasets because they are external memory based. The most dominant costs for existing LSH techniques are the algorithm time and the index I/Os required to find candidate points. Existing works do not compare both of these dominant costs in their evaluation. In this experimental survey paper, we show the impact of both these costs on the overall performance of the LSH technique. We compare three state-of-the-art techniques on four real-world datasets, and show that, in contrast to recent works, C2LSH is still the state-of-the-art algorithm in terms of performance while achieving similar accuracy as its recent competitors. 

\end{abstract}

\section{Introduction}
Many large multimedia retrieval applications require efficient processing of nearest neighbor queries in high-dimensional spaces. Exact tree-based indexing structures, such as KD-tree, SR-tree, etc., work well for low-dimensional spaces ($<10$) but suffer from the notorious \textit{curse of dimensionality} for high-dimensional spaces. They are often outperformed by brute-force linear scans \cite{Chavez:2001:SMS:502807.502808}. One solution to this problem is to search for \textit{good enough} approximate results instead. Approximate techniques sacrifice some accuracy for a significant improvement in the overall processing time. In many applications where 100\% is not needed, this tradeoff is very useful in saving time. The goal of the approximate version of the nearest neighbor problem, also called \textit{c-approximate Nearest Neighbor search}, is to return points that are within $c*R$ distance from the query point. Here, $c>1$ is a user-defined approximation ratio and $R$ denotes the distance of the query point and its nearest neighbor.

\subsection{Locality Sensitive Hashing}
Locality Sensitive Hashing (LSH) \cite{Gionis:1999:SSH:645925.671516} is one of the most popular techniques for finding approximate nearest neighbors in high-dimensional spaces. LSH was first introduced in \cite{Gionis:1999:SSH:645925.671516} for the Hamming distance, but was later extended to several distances, such as the popular Euclidean distance \cite{Datar:2004:LHS:997817.997857}. LSH uses \textit{random} hash projections to map the original high-dimensional space to the projected low-dimensional space. The main idea behind LSH is that nearby points in the original high-dimensional space will map to similar hash buckets in the low-dimensional space with a higher probability than mapping to dissimilar or far away points to the same buckets. Since LSH was first proposed in \cite{Gionis:1999:SSH:645925.671516}, there have been several works that have focused on improving the search accuracy and/or performance \cite{Bawa:2005:LFS:1060745.1060840,Gan:2012:LHS:2213836.2213898,Huang:2015:QLH:2850469.2850470,Liu:2014:SEI:2732939.2732947,Lv:2007:MLE:1325851.1325958,Tao:2010:EAN:1806907.1806912,Liu:2019,Tobias:2019}.

\subsection{Motivation for using LSH}
\label{sec:motLSH}
Locality Sensitive Hashing (LSH) is known for two main advantages: its sub-linear query performance (in terms of the data size)
and theoretical guarantees on the query accuracy. Additionally, LSH uses random hash functions which are data-independent (i.e. data properties such as data distribution are not needed to generate these random hash functions). 
Since LSH uses random hash functions, the generation of these hash functions is a simple process that takes negligible time.
Additionally, the data distribution does not affect the generation of these hash functions. Hence, in applications where data is changing or where newer data is coming in, these hash functions do not require any change during runtime. 
While the original LSH index structure suffered from large index sizes (in order to obtain a high query accuracy) \cite{Bawa:2005:LFS:1060745.1060840,Lv:2007:MLE:1325851.1325958}, state-of-the-art LSH techniques \cite{Gan:2012:LHS:2213836.2213898,Huang:2015:QLH:2850469.2850470} have alleviated this issue by using advanced methods such as \textit{Collision Counting} and \textit{Virtual Rehashing}. In addition to their fast index maintenance, fast query performance, and theoretical guarantees on the query accuracy, LSH algorithms are easy to implement as external memory-based algorithms, and hence are more scalable than in-memory algorithms (such as graph-based ANN algorithms) \cite{Liu:2019}.

\subsection{Motivation of our Survey}

Locality Sensitive Hashing techniques have two dominant costs for finding nearest neighbors: 1) cost of reading the index files from the external memory to the main memory (which we call \textit{Index I/Os}), and 2) cost of finding candidates and removing false positives (which we call \textit{Algorithm time}). As mentioned in Section \ref{sec:motLSH}, one of the benefits of LSH is that it is a scalable algorithm. Some of the existing LSH techniques (e.g. C2LSH \cite{Gan:2012:LHS:2213836.2213898} and QALSH \cite{Huang:2015:QLH:2850469.2850470}) are not entirely external memory-based (i.e. even though the indexes are stored on the disk, their implementations require the entire data and indexes should fit into the main memory during the index creation phase). Thus, existing works (such as \cite{Arora:2018:HPS:3204028.3228393}) do not compare their results with C2LSH and QALSH on large datasets since they do not fit in the main memory. Additionally, some recent works (such as \cite{Liu:2019}) only compare the \textit{Index I/Os} without comparing the important \textit{Algorithm time}. This leads to other recent papers (such as \cite{ANNSurvey:8681160,9101839,PM-LSH/3377369.3377374}) to unfairly compare their \textit{Algorithm time} with QALSH or I-LSH \cite{Liu:2019} since they are deemed as the state-of-the-art LSH techniques. 

\subsection{Contributions of this Survey paper}
\label{sec:contrib}
In this paper, we carefully present a detailed experimental analysis on three state-of-the-art LSH algorithms, C2LSH \cite{Gan:2012:LHS:2213836.2213898}, QALSH \cite{Huang:2015:QLH:2850469.2850470}, and I-LSH \cite{Liu:2019}. Our contributions are as follows:
\begin{itemize}
	\item We modify the implementations of C2LSH and QALSH to create fully external memory-based implementations such that the entire dataset and/or the entire index do not need to be in the main memory for the algorithms to work during index generation or query processing.\footnote{These implementations will be made public.}
	\item We show the importance of experimentally analyzing and comparing the \textit{Index I/Os} and \textit{Algorithm time} of all algorithms. 
	\item We compare these three algorithms on real datasets with different characteristics under differing system parameters.
\end{itemize}

\noindent To the best of our knowledge, we are the first work to present a detailed analysis of these three state-of-the-art LSH techniques, namely C2LSH \cite{Gan:2012:LHS:2213836.2213898}, QALSH \cite{Huang:2015:QLH:2850469.2850470}, and I-LSH \cite{Liu:2019}.

\section{Related Work}
\label{sec:relWork}
Nearest Neighbor problem is an important problem for multimedia applications in many diverse domains such as multimedia retrieval, image processing, machine learning, etc. Since tree-based index structures can be outperformed by a linear scan, due to the \textit{curse of dimensionality}, in high-dimensional spaces, approximate techniques are preferred due to their fast performance at the expense of some accuracy.
Due to the importance of the nearest neighbor problem in various domains, several diverse techniques have been proposed by researchers. These techniques can be broadly classified into three main categories: Hashing-based methods, Partition-based methods, and Graph-based methods.\footnote{We refer the reader to a recent survey \cite{ANNSurvey:8681160} for an in-depth survey on these categories.} Hashing-based methods can be further classified into learning-based hashing techniques and random hashing techniques. The benefit of random hashing techniques, such as Locality Sensitive Hashing \cite{Gionis:1999:SSH:645925.671516}, are that they are easy to construct, no need for training data, and easy to maintain and update. Additionally, LSH provides a sub-linear (in terms of the data size) query performance and theoretical guarantees on the query accuracy. \\
\textbf{Locality Sensitive Hashing and its variants: }The main idea of Locality Sensitive Hashing is to create random projections and hash data points in these random projections such that nearby data points in the original high-dimensional space will be mapped to the same hash bucket with a high probability (and conversely, data points that are far apart from each other in the original high-dimensional space will be mapped to the same hash bucket with a low probability). It was originally proposed in \cite{Gionis:1999:SSH:645925.671516} for the Hamming distance and then later extended to the popular Euclidean distance \cite{Datar:2004:LHS:997817.997857}. In this original work on Euclidean distance (E2LSH), instead of a single hash function (or a projection), a hash table consisted of several hash functions (represented by Compound Hash Keys) in order to reduce false positives. But this also generated false negatives. Hence several hash tables had to be used to reduce the number of false positives and false negatives, while keeping the accuracy of the query high. The main drawbacks of this approach were the size of the index structure (since large number of hash tables were required to return the desired number of results with a high accuracy) and the need to determine the width of the hash bucket during index creation (a larger width returned enough results but also with a potential of too many false positives, whereas a smaller width had a potential of misses resulting in insufficient results). This user-defined width, which was mainly dependent on the data distribution, had to be often determined through a trial and error process. \\
LSH-Forest \cite{Bawa:2005:LFS:1060745.1060840} was proposed where the compound hash-keys were hierarchically stored such that the algorithm could stop at a higher level in the tree if more results were needed. In Multi-probe LSH \cite{Lv:2007:MLE:1325851.1325958}, the authors proposed a technique to probe into neighboring buckets when more results were needed. The intuition is that neighboring buckets are more likely to contain nearby points. Hence, if the bucket width was underestimated (which is better than overestimation which can lead to significant wasteful processing), neighboring buckets were probed to find the desired number of results. \\
Later, C2LSH \cite{Gan:2012:LHS:2213836.2213898} introduced two main concepts of \textit{Collision Counting} and \textit{Virtual Rehashing} that solved the two main drawbacks of E2LSH \cite{Datar:2004:LHS:997817.997857}. In C2LSH, the authors proposed to create $m$ base hash functions and choose candidate points based on how many times a data point collides with the query point (and hence instead of creating several hash tables of several hash functions, only 1 table of $m$ base hash functions is needed), which reduced the size of the index structure. Additionally, in \textit{Virtual Rehashing}, the neighboring buckets in each hash function are read incrementally when sufficient number of results are not found. \\ In SK-LSH \cite{Liu:2014:SEI:2732939.2732947}, the authors propose a linear ordering on the Compound Hash Keys (using a space-filling curve) such that nearby Compound Hash Keys are stored on the same (or nearby) page on the disk, thus reducing the total number of I/Os. The design of SK-LSH is still build on the original E2LSH, and hence suffers from the parameter tuning problem, where the user is expected to enter important parameters such as number of hash functions and the radius at which $k$ results will be found. Wrong choice of parameters can negatively affect the accuracy and efficiency of the algorithm. \\ QALSH \cite{Huang:2015:QLH:2850469.2850470} was later proposed that built query-aware hash functions such that the hash value of the query point is considered as the anchor bucket during query processing and this idea would solve the issue when close points to a query were partitioned into different buckets when query was near the bucket boundaries. Additionally, B+trees are built on each hash function for efficient lookups into neighboring buckets (which translate to range queries). QALSH utilizes the concepts of \textit{Collision Counting} and \textit{Virtual Rehashing}. \\
HD-Index \cite{Arora:2018:HPS:3204028.3228393} was introduced which generated Hilbert keys of the dataset points and also stored the distances of the points to each other to efficiently prune the results based on distance filters. HD-Index stores the Hilbert keys using modified B+-trees, called RDB-trees. Due to the reliance on space-filling curves (Hilbert curves) and B+-trees, HD-Index cannot scale for moderately high-dimensional datasets \cite{Arora:2018:HPS:3204028.3228393}. \\ SRS \cite{Sun:2014:SSC:2735461.2735462} uses the Euclidean distance between two points in the projected space to estimate their distance in the original space. In order to find the next nearest neighbor in the projected space, SRS uses an R-tree to index the points in the projected space. This incremental finding of the NN is similar to I-LSH. The main goal of SRS is to introduce a very lightweight index structure to solve the ANN problem. SRS is shown to suffer from memory leaks and slow running times as compared with C2LSH \cite{Arora:2018:HPS:3204028.3228393}, and hence not included in our work. \\
Recently, I-LSH \cite{Liu:2019}, which is considered to be the state-of-the-art LSH technique \cite{9101839}, was proposed to improve the Virtual Rehashing process of QALSH (where the range of the lookups are incremented exponentially). In I-LSH, the authors propose to increase the range of the lookups based on the distance to the nearest point (in the projected space) instead of increasing the range exponentially. While this strategy results in less disk I/Os, it also leads to high disk seeks (random I/Os) and algorithm time as we show in Section \ref{sec:exp}. \\
Very recently, PM-LSH \cite{PM-LSH/3377369.3377374} was proposed where the idea was to estimate the Euclidean distance based on a tunable confidence interval value such that the overall query processing time is reduced.\footnote{The code of PM-LSH was not released before the submission date of SISAP.}

\section{Background and Preliminaries}

In this section, we describe the key concepts behind LSH. We primarily use the terminologies and formulations introduced in E2LSH \cite{Datar:2004:LHS:997817.997857} and C2LSH \cite{Gan:2012:LHS:2213836.2213898}.

\noindent\textbf{Hash Functions: }A hash function family $H$ is ($R$, $cR$, $p_1$, $p_2$)-sensitive if it satisfies the following conditions for any two points $x$ and $y$
in a $d$-dimensional dataset $D \subset \mathbb{R}^d$:

\begin{itemize}
	\item if $|x - y| \leq R$, then $Pr[h(x) = h(y)] \geq p_1$, and
	\item if $|x - y| > cR$, then $Pr[h(x) = h(y)] \leq p_2$
\end{itemize}

\noindent Here, $p_1$ and $p_2$ are probabilities and $c$ is an approximation ratio. LSH requires that $c > 1$ and $p_1 > p_2$. 
The above definition states that the two points $x$ and
$y$ are hashed to the same bucket with a very high probability $\geq p_1$ if they are close to each other (i.e. the distance between the two points is less than or equal to $R$), and if they are
not close to each other (i.e. the distance between the two points is greater than $cR$), then they will be hashed to the same bucket with a low probability $\leq p_2$. 
In the original LSH scheme for Euclidean distance, each hash function is defined as $h_{\vec{a},b} (x) = \left\lfloor{\frac{\vec{a}.x + b}{w}}\right\rfloor,$ where $\vec{a}$ is a $d$-dimensional random vector with entries chosen
independently from the standard normal distribution $N(0,1)$ and $b$ is a real number chosen uniformly from $[0, w)$, such that $w$ is the width of the hash bucket \cite{Datar:2004:LHS:997817.997857}.  
This leads to the following collision probability function~\cite{Datar:2004:LHS:997817.997857}, which states that if $||x,y||=r$, then the probability that $x$ and $y$ map to the same hash bucket for a given hash function $h_{\vec{a},b} (x)$ is:
$P(r) = \int_{0}^{w}{\frac{1}{r} \frac{2}{\sqrt{2\pi}} e^{\frac{-t^2}{2r^2}}(1-\frac{t}{w}) dt}$. 
Here, the collision probability $P(r)$ is decreasing on $r$ for a given $w$. 
For a $t$, which is the largest absolute value of a coordinate of point in $D$, and for every $b$ uniformly drawn from the interval $[0,c^{\lceil\log_c td\rceil}w^2]$ and $R=c^n$ for some $n\leq \lceil\log_c td\rceil$ we have that 
$h^R(x)=\left\lfloor\frac{h_{\vec{a},b} (x)}{R}\right\rfloor$
is $(R,cR,p_1,p_2)$-sensitive, where $p_1=p(1)$ and $p_2=p(c)$ \cite{Gan:2012:LHS:2213836.2213898}.

\section{State-of-the-art Techniques}
\label{sec:state}

In Section \ref{sec:relWork}, we explained the benefits and drawbacks of different LSH techniques. In this paper, we will experimentally analyze the three state-of-the-art external memory-based LSH techniques, C2LSH \cite{Gan:2012:LHS:2213836.2213898}, QALSH \cite{Huang:2015:QLH:2850469.2850470}, and I-LSH \cite{Liu:2019}. In this section, we will introduce the concepts introduced by these techniques. 
C2LSH \cite{Gan:2012:LHS:2213836.2213898} introduced the concepts of \textit{Collision Counting} and \textit{Virtual Rehashing}. In \cite{Gan:2012:LHS:2213836.2213898}, authors theoretically show that two close points $x$ and $y$ collide in at least $l$ hash layers with a probability $1-\delta$, when the total number,
$m$, of hash layers are equal to:
$m = \ceil[\big]{\frac{\ln(\frac{1}{\delta})}{2(p_1-p_2)^2}(1+z)^2}$.
Here, $z=\sqrt{\ln({\frac{2}{\beta}})/\ln({\frac{1}{\delta}}})$, where $\beta$
is the allowed false positive percentage (i.e. the allowed number of points whose distance with a query point is greater than $cR$). C2LSH sets $\beta=\frac{100}{n}$, where $n$ is the cardinality of the dataset. Further, only those points that collide at least $l$ times, 
where $l$ is the collision count threshold, which is calculated as following: $l = \lceil\alpha \times m\rceil$, where the collision threshold percentage, $\alpha$, is 
$\alpha = \frac{zp_1 + p_2}{1+z}$.
C2LSH creates only one hash function per hash table, and hence the number of hash functions are equal to the number of hash table. 

Instead of assuming a \textit{magic} radius (which traditional LSH methods did), C2LSH sets the initial radius $R$ to 1. It is possible that with $R=1$, there are not enough results for a top-$k$ query to be returned. C2LSH increases the radius of the query in the following sequence: $R = 1, c, c^2, c^3...$. If at \textit{level-R}, enough candidates are not found, the radius is increased until enough query results are found. This exponential expansion process is called \textit{Virtual Rehashing}. 

Moreover, C2LSH uses two terminating conditions to stop the algorithm when the conditions are met. These conditions specify that 1) at the end of each virtual rehashing at least $k$ candidates should have been found whose Euclidean distance to the query are less than or equal to $cR$, and 2) at any point, $k+\beta n$ candidates are found.

QALSH introduces \textit{query-aware} hash functions $h_{\vec{a}} (x) = {\vec{a}.x}$. For a query $q$, once the query projection is found by computing $h_{\vec{a}} (q)$, QALSH uses the query as the ``anchor" to find the anchor bucket with width $w$ with the interval $|h_{\vec{a}} (q) - \frac{w}{2}, h_{\vec{a}} (q) + \frac{w}{2}|$. If the projected location for a point $x$ falls in the same anchor bucket as $q$, i.e., $|h_{\vec{a}} (o) -h_{\vec{a}} (q)| \leq \frac{w}{2}$, then QALSH considers that $o$ has collided with $q$ under $h_{\vec{a}}$. QALSH \cite{Huang:2015:QLH:2850469.2850470} also utilizes these concepts of Collision Counting and Virtual Rehashing to build \textit{query-aware} hash functions. Another main difference of QALSH is that it uses B+-trees to represent the hash tables. An exponential expansion in each hash table is thus the same as a range query on a B+-tree. By using \textit{query-aware} hash functions and B+-trees, QALSH improves the theoretical bounds by reducing the total number of hash functions required to satisfy the quality guarantee. Additionally, QALSH can work for any approximation ratio, $c$, greater than 1, while C2LSH can only work for $c \geq 2$. While the reduction in number of hash functions generates a smaller index, the overhead of using B+-trees makes QALSH much slower as we experimentally show in Section \ref{sec:exp}. 

I-LSH \cite{Liu:2019} uses the query-aware hash functions (that are proposed by QALSH) and proposes an incremental expansion strategy to reduce the overall index I/Os. In order to do that, i-LSH finds the next closest point in each projection. While this process leads to less overall index I/Os, it still requires disk seeks and (as we show in Section \ref{sec:exp}) the algorithm overhead is far more than the savings in the disk I/Os. 

\section{Experimental Analysis}
\label{sec:exp}

In this section, we first explain our carefully designed experimental evaluation plan. We experimentally analyze C2LSH, QALSH, and I-LSH on different datasets and report the results for varying criteria. All experiments were run on the nodes of the Bigdat cluster \footnote{Supported by NSF Award \#1337884} with the following specifications: two Intel Xeon E5-2695, 256GB RAM, and CentOS 6.5 operating system. All codes were written in C++11 and compiled with gcc v4.7.2 with the -O3 optimization flag. As mentioned in Section \ref{sec:contrib}, we extend the implementations of C2LSH and QALSH to be completely external-memory based implementations (i.e. the entire dataset or the index files are not needed to be in the main memory in order to construct the LSH indexes). 

\subsection{Datasets}
We use the following six diverse high-dimensional datasets with varying cardinality and dimensionality: 

\begin{itemize}
	\item \textbf{P53}\cite{danziger2009predicting} consists of $31,002$ 5409-dimensional points which are generated based on the biophysical features of mutant p53 proteins and can be used to predict p53 transcriptional activity. The values of this dataset are normalized between zero and $10,000$ and duplicate rows are removed.
		
	\item \textbf{LabelMe}\cite{russell2008labelme} consists of $181,093$ 512-dimensional points which were generated by running the GIST feature extraction algorithm on $30369$ annotated images belonging to $183$ categories. There are no duplicates in the dataset and values range between zero and $58104$.
	
	\item \textbf{Sift1M}\cite{jegou2010product} consists of $1,000,000$ 128-dimensional points that were created by running the SIFT feature extraction algorithm on real images. The values of this dataset are integers between zero and $218$.
	
	\item \textbf{Deep1M} consists of $1,000,000$ 96-dimensional points sampled from the Deep1B dataset introduced in \cite{babenko2016efficient}. These points are extracted from the last layers of convolutional neural networks for images.
	
	\item \textbf{Mnist8M}\cite{loosli2007training} This dataset, also known as the InfiMNIST dataset, contains $8,100,000$ 784-dimensional points that represent images of the digits 0 to 9 which are grayscale and of size 28 $\times$ 28.
	
	\item \textbf{Tiny80M}\cite{torralba200880} This dataset contains $79,302,017$ 384-dimensional points generated using Gist feature extraction algorithm on $80$ million $32 \times 32$ colored images and its values are normalized between zero and $255$.
\end{itemize}

All datasets are normalized to contain only integers since C2LSH requires the data format to be only integers \cite{Gan:2012:LHS:2213836.2213898}.

\subsection{Evaluation Criteria and Parameters}

The goal of our paper is to present a detailed analysis of the performance of the state-of-the-art LSH techniques. We also compare the accuracy of these algorithms. We randomly choose 50 queries from the dataset and report the average of the results of these 50 queries. We used the same parameters suggested in their papers ($w=2.781$ for QALSH and $w=2.184$ for C2LSH). We choose $\delta=0.1$ and $c=2$ (since C2LSH cannot give guarantees for $c<2$). 
Since our goal is to present the performance analysis of query processing, we do not compare the index time and index construction time of these three algorithms. Since I-LSH uses the same hash functions as QALSH, their index size and index construction time are the same. \cite{Huang:2015:QLH:2850469.2850470} shows the difference between these two criteria for C2LSH and QALSH for different datasets, and hence we avoid it in this paper. 

After careful analysis of performance of LSH techniques, we present the following breakdown of the query processing time ($QPT$): 
\begin{itemize}
	\item \textbf{Index Read Cost: }LSH techniques need to read index files (from the external memory) in order to find the candidates. This dominant cost of reading index files can be further broken down into the number of disk seeks (i.e. random I/Os) and the total amount of data read. Following \cite{Liu:2019}, we also consider the number of disk seeks and amount read in our cost formulation. 
	\item \textbf{Algorithm Time: }Another dominant cost in LSH processing is the processing of index files once they are read into the main memory. LSH techniques need to find points that are considered as candidates. Techniques such as Collision Counting (explained in \ref{sec:state}) are included in this cost. 
	\item \textbf{False Positive Removal Cost: }Once a point is deemed as a candidate, the LSH technique brings the actual data point (resulting in a random seek) into the main memory to calculate the Euclidean distance with the query point. Since the state-of-the-art LSH techniques have an upper bound of the number of candidates that are generated (which is set to $k+100$), this cost is negligible as compared to the previous two costs. 
\end{itemize}

It is well-known that random I/Os are much more expensive than sequential I/Os \cite{10.1007/s00778-017-0480-7}. Additionally, the difference in the cost changes significantly depending on whether the external storage medium is an HDD or an SSD. The difference in the costs of random I/Os and sequential I/Os is significantly more in HDDs than in SSDs (mainly because random disk seeks are faster in SSDs than HDDs) \cite{10.1145/3209900.3209903}. We noticed that the number of disk seeks are significantly different in these state-of-the-art LSH techniques due to their strategy in finding neighboring points in projected spaces. Hence, we model the overall \textit{Query Processing Time} (QPT) for both HDDs and SSDs. For an HDD, we use the reported benchmarks for Seagate Barracuda HDD with 7200 RPM and 1TB: average disk seek requires 8.5 ms and the average data read rate is 0.156 MB/ms \cite{SeagateHDD}. Similarly, for an SSD, we use the reported benchmarks for the Seagate Barracuda 120 SSD with 1TB storage: average disk seek requires 0.01 ms and the average data read rate is 0.56 MB/ms \cite{SeagateSSD}. 

We use the same accuracy measure, the overall ratio, used in several prior works \cite{Gan:2012:LHS:2213836.2213898,Huang:2015:QLH:2850469.2850470,Liu:2014:SEI:2732939.2732947,Liu:2019}: $\frac{1}{k}\sum_{i=1}^{k}\frac{||o_i, q||}{||o_i^*,q||}$. Here, $o_i$ is the $i$th point returned by the technique and $o_i^*$ is the true $i$th nearest point from $q$ (ground truth). Ratio of 1 means the returned results have the same distance from the query as the ground truth. The closer the ratio is to 1, the higher is the accuracy of the LSH technique.

\begin{figure*}[!h]
	\centering
	\begin{subfigure}[b]{0.31\textwidth}
		\centering
		{\includegraphics[width=\linewidth]{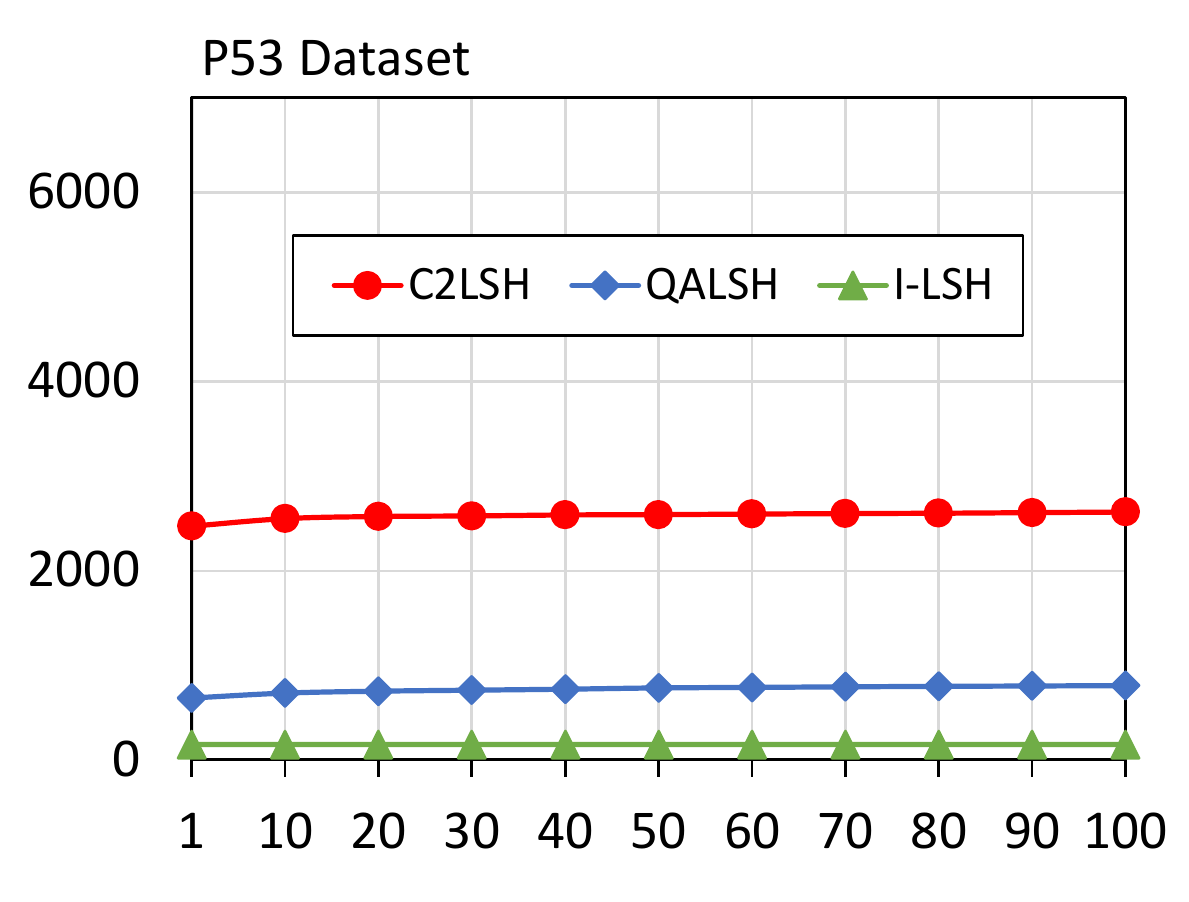}}
	\end{subfigure}\quad
	\begin{subfigure}[b]{0.31\textwidth}
		\centering
		{\includegraphics[width=\linewidth]{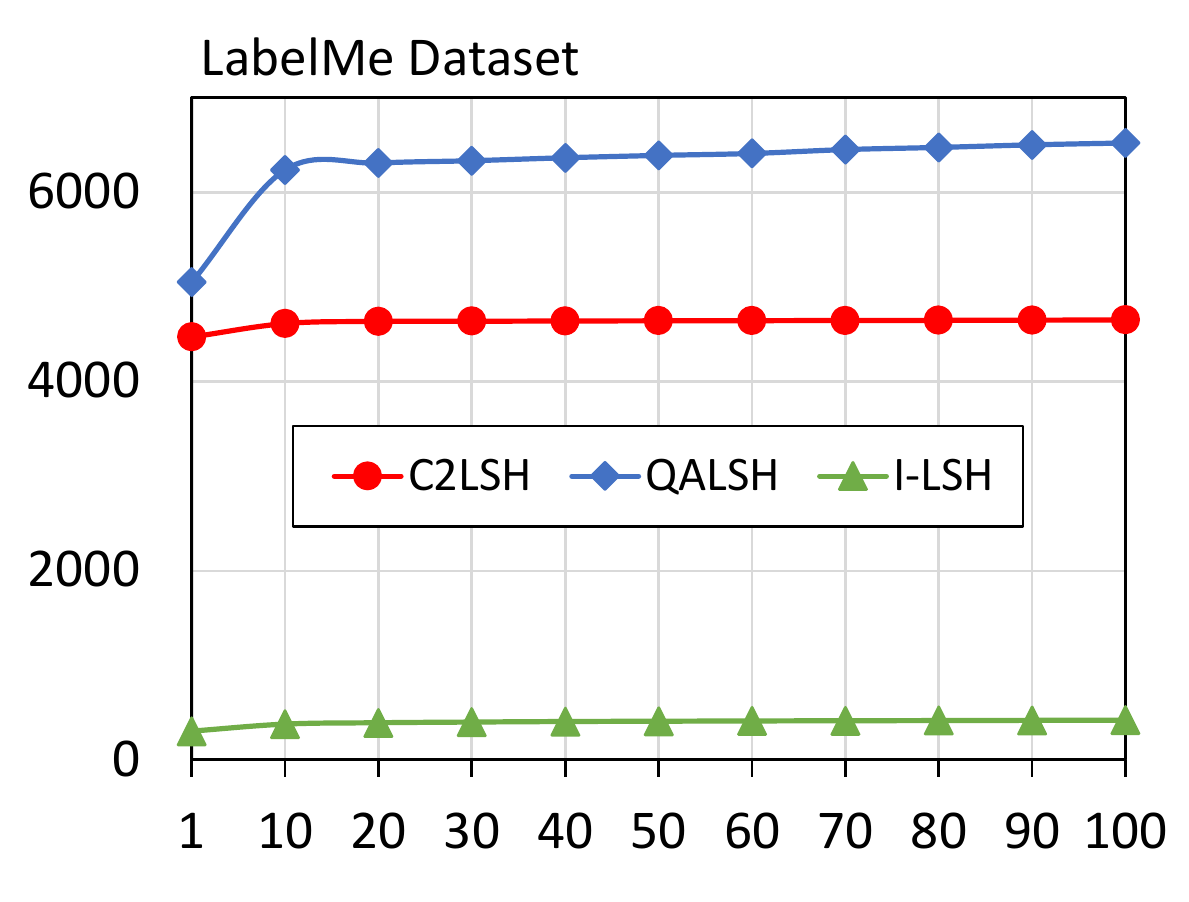}}
	\end{subfigure}\quad
	\begin{subfigure}[b]{0.31\textwidth}
		\centering
		{\includegraphics[width=\linewidth]{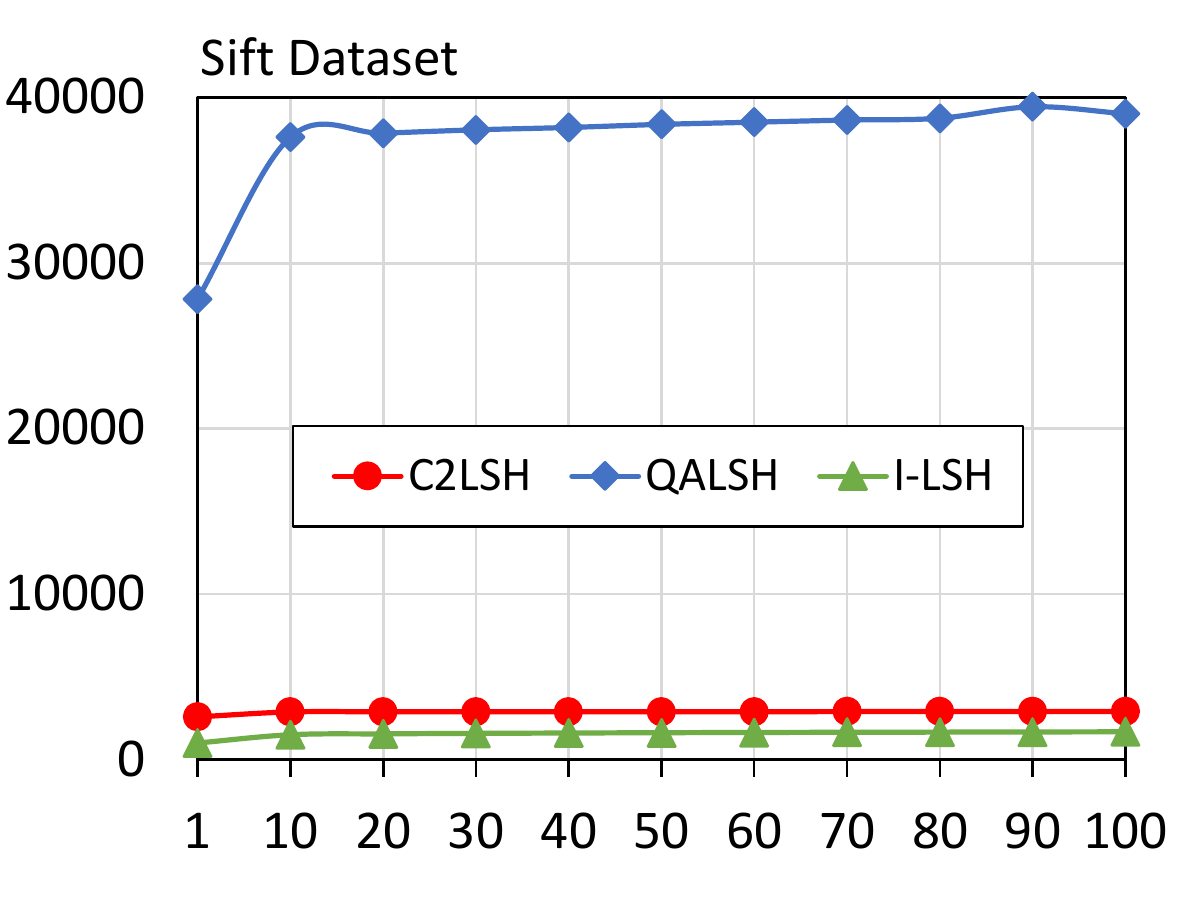}}
	\end{subfigure} \\
	
	\begin{subfigure}[b]{0.31\textwidth}
		\centering
		{\includegraphics[width=\linewidth]{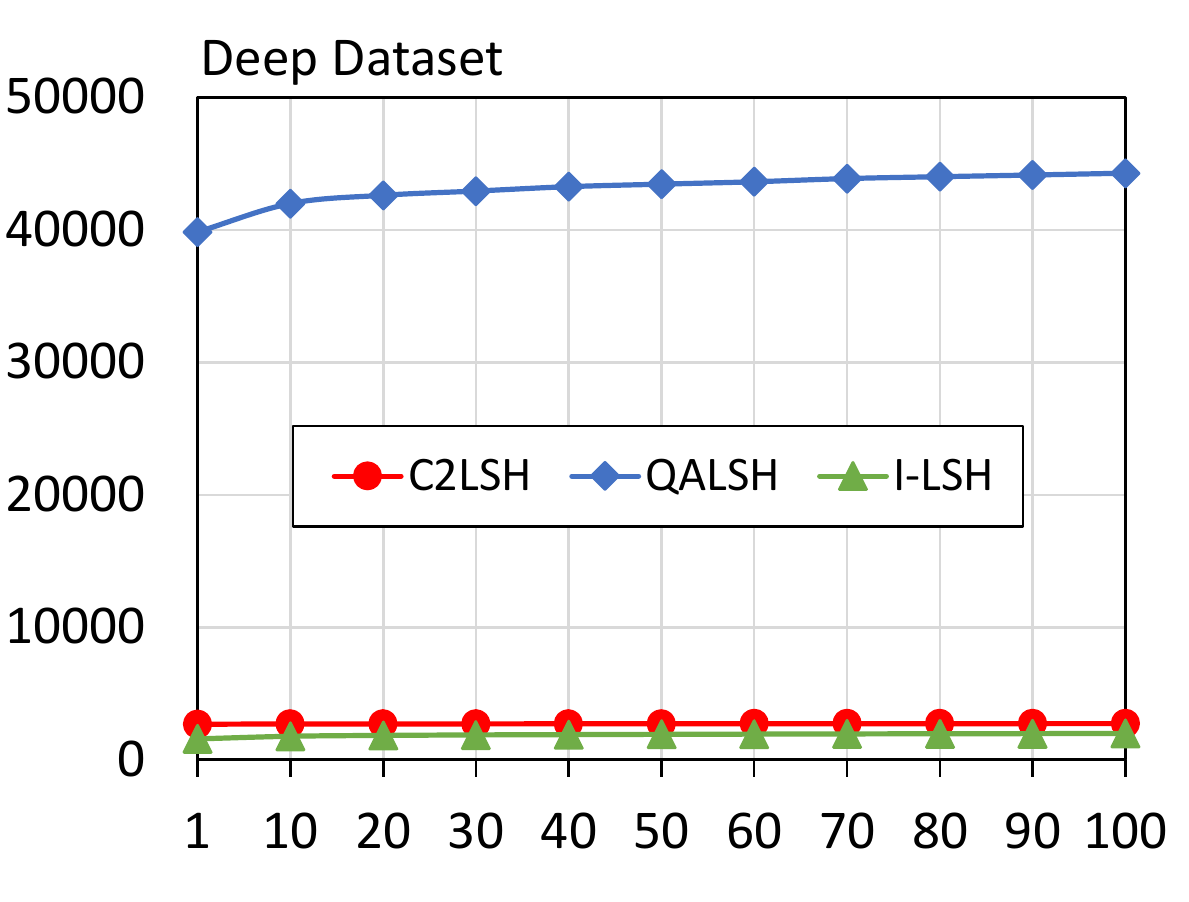}}
	\end{subfigure}\quad
	\begin{subfigure}[b]{0.31\textwidth}
		\centering
		{\includegraphics[width=\linewidth]{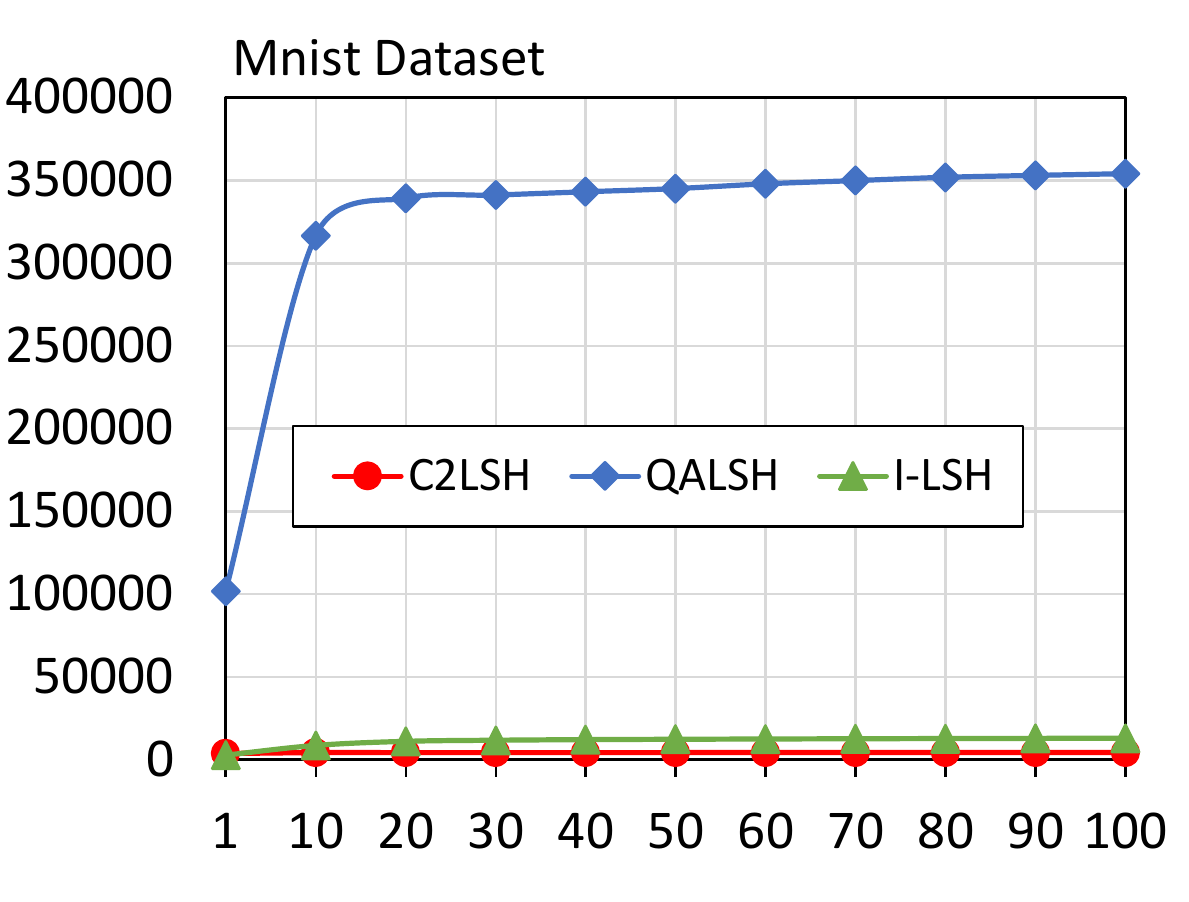}}
	\end{subfigure}\quad
	\begin{subfigure}[b]{0.31\textwidth}
		\centering
		{\includegraphics[width=\linewidth]{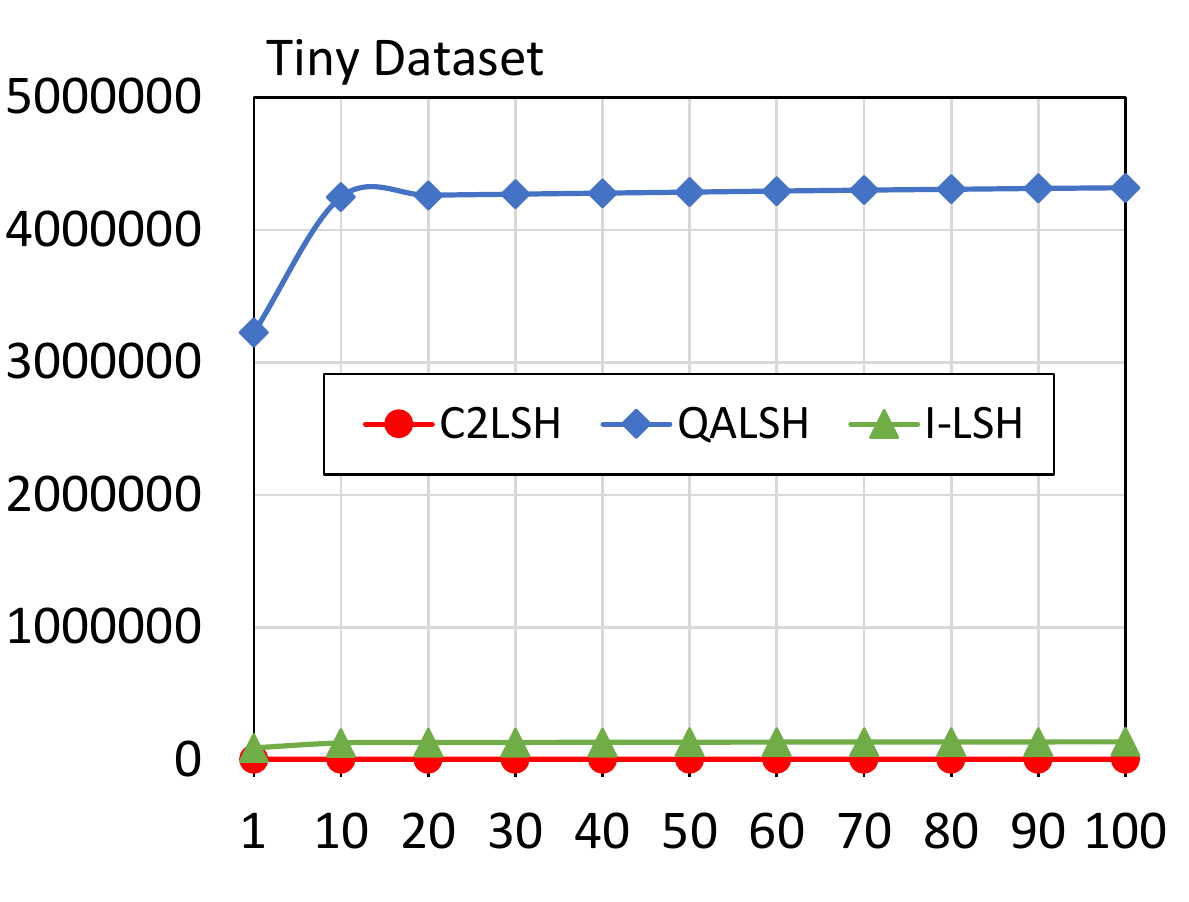}}
	\end{subfigure}
	
	\caption{Number of Disk Seeks (Y axis) for different $k$ (X Axis) on 6 datasets}
	\label{fig:expdiskseek}
\end{figure*}

\begin{figure*}[!h]
	\centering
	\begin{subfigure}[b]{0.31\textwidth}
		\centering
		{\includegraphics[width=\linewidth]{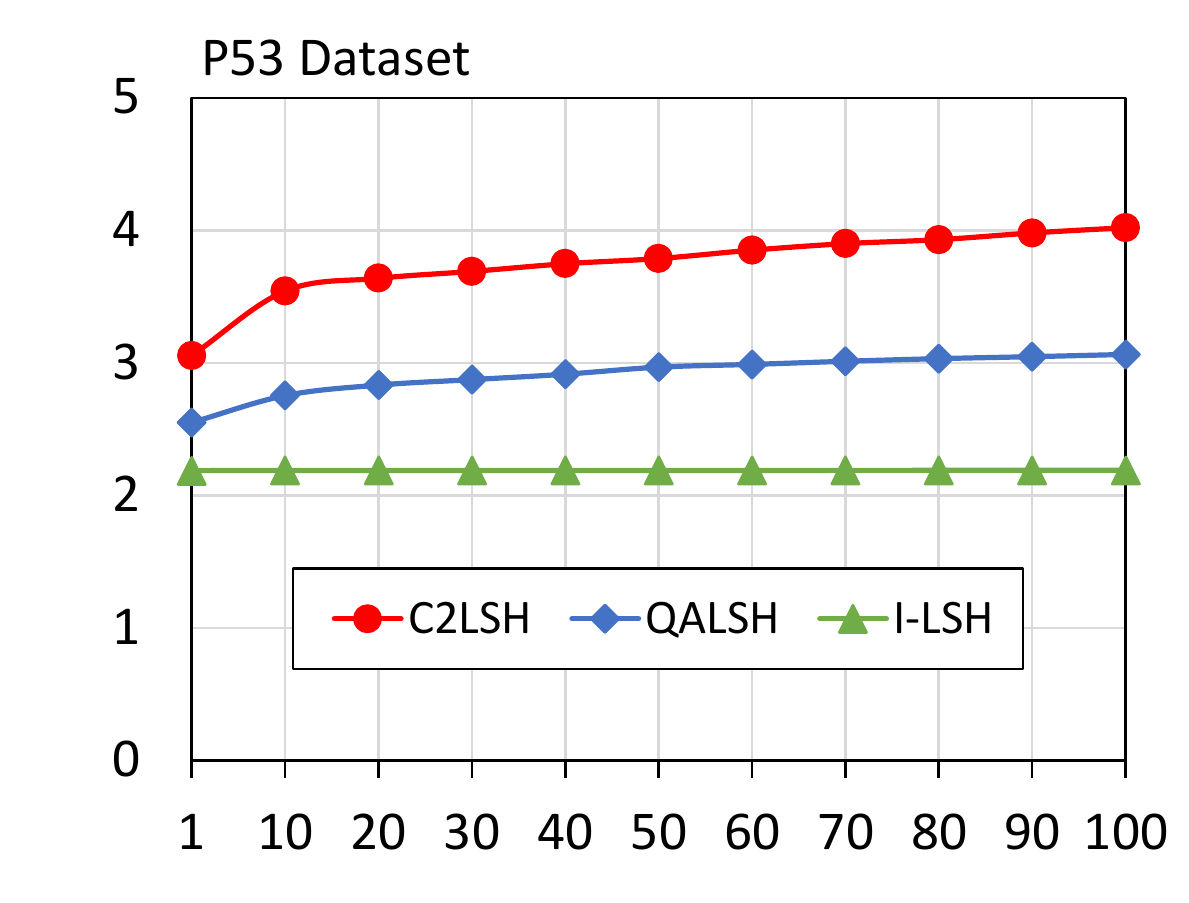}}
	\end{subfigure}\quad
	\begin{subfigure}[b]{0.31\textwidth}
		\centering
		{\includegraphics[width=\linewidth]{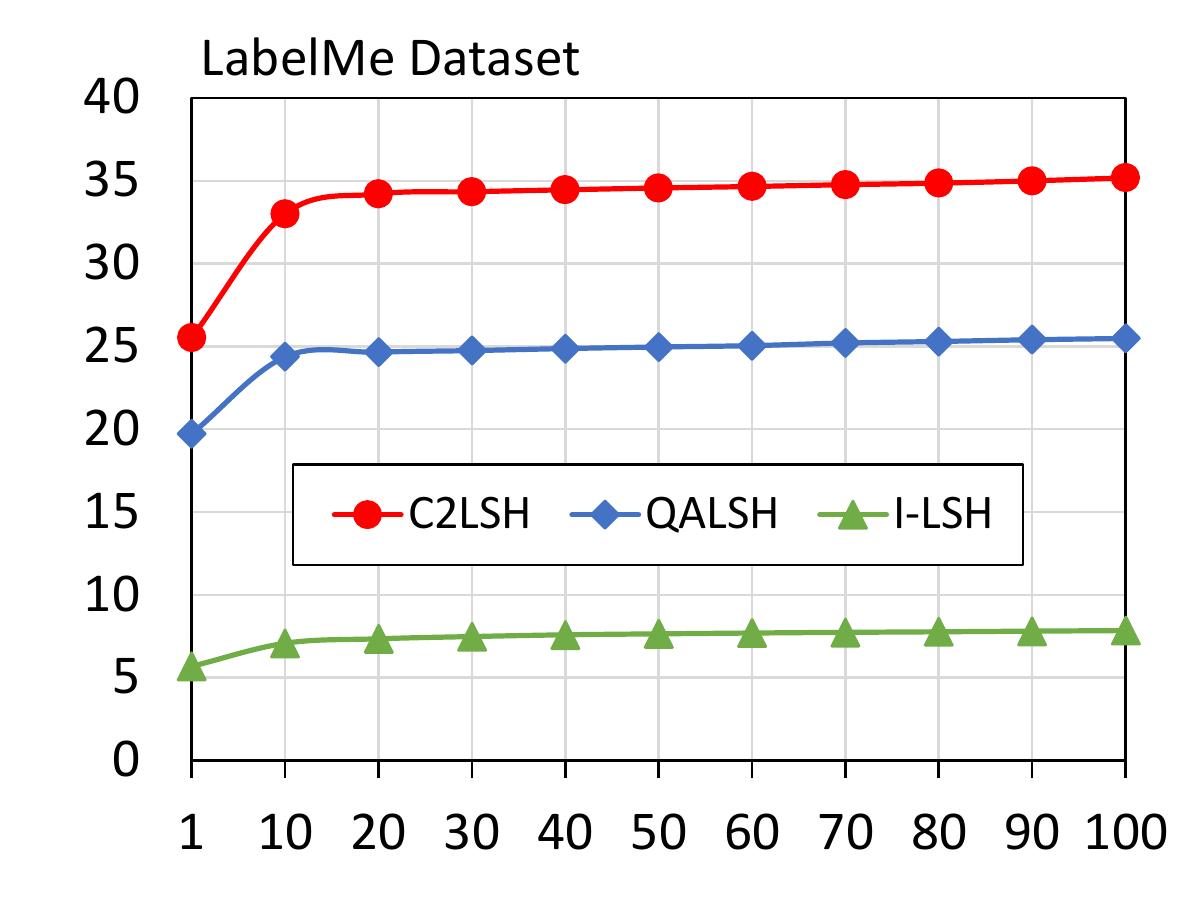}}
	\end{subfigure}\quad
	\begin{subfigure}[b]{0.31\textwidth}
		\centering
		{\includegraphics[width=\linewidth]{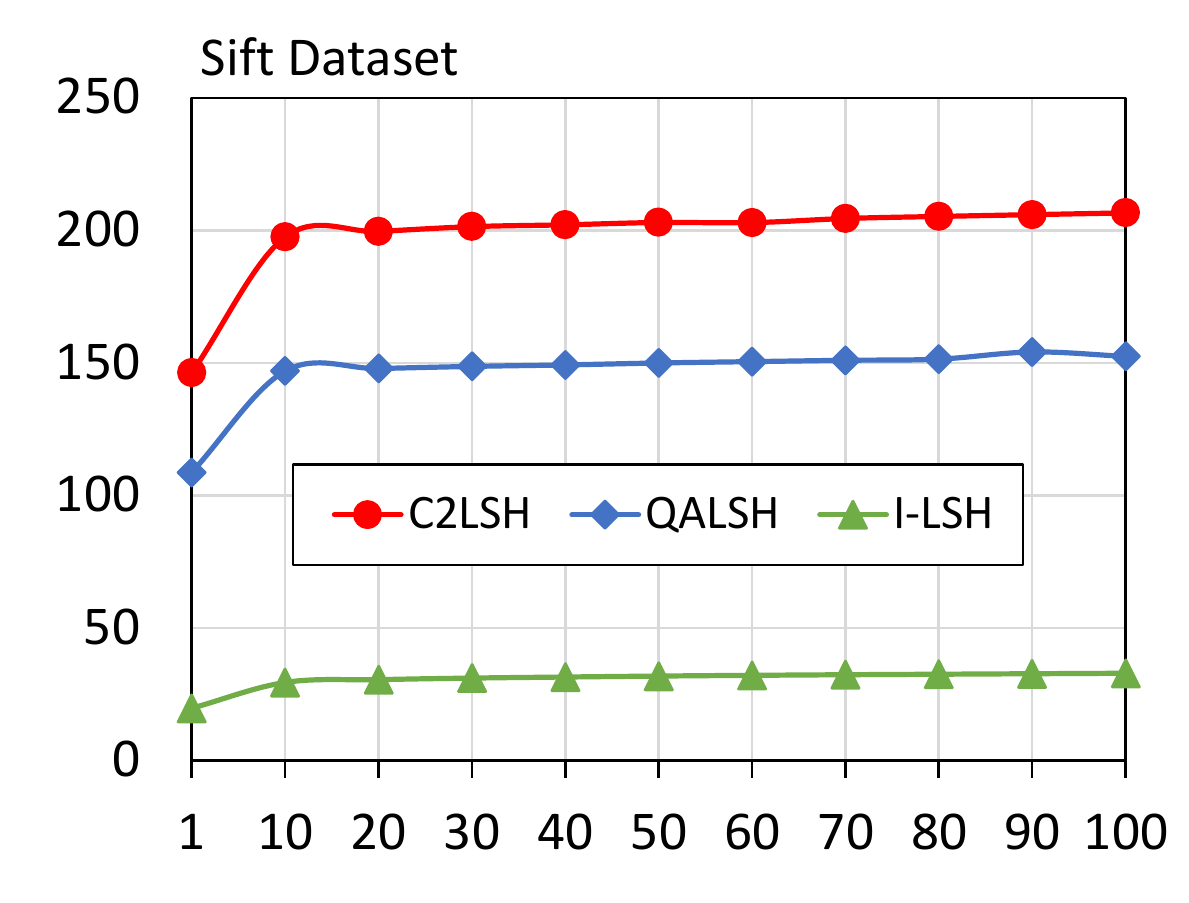}}
	\end{subfigure} \\
	
	\begin{subfigure}[b]{0.31\textwidth}
		\centering
		{\includegraphics[width=\linewidth]{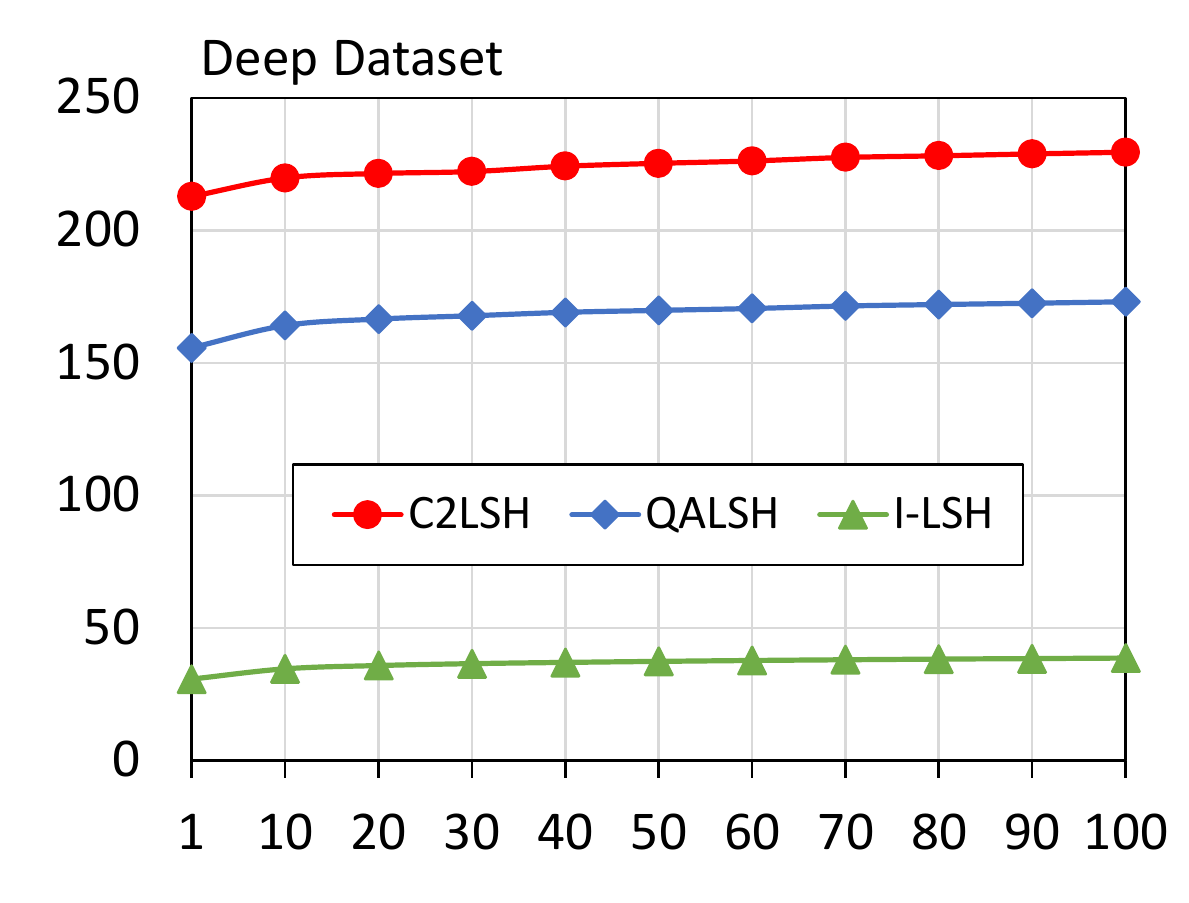}}
	\end{subfigure}\quad
	\begin{subfigure}[b]{0.31\textwidth}
		\centering
		{\includegraphics[width=\linewidth]{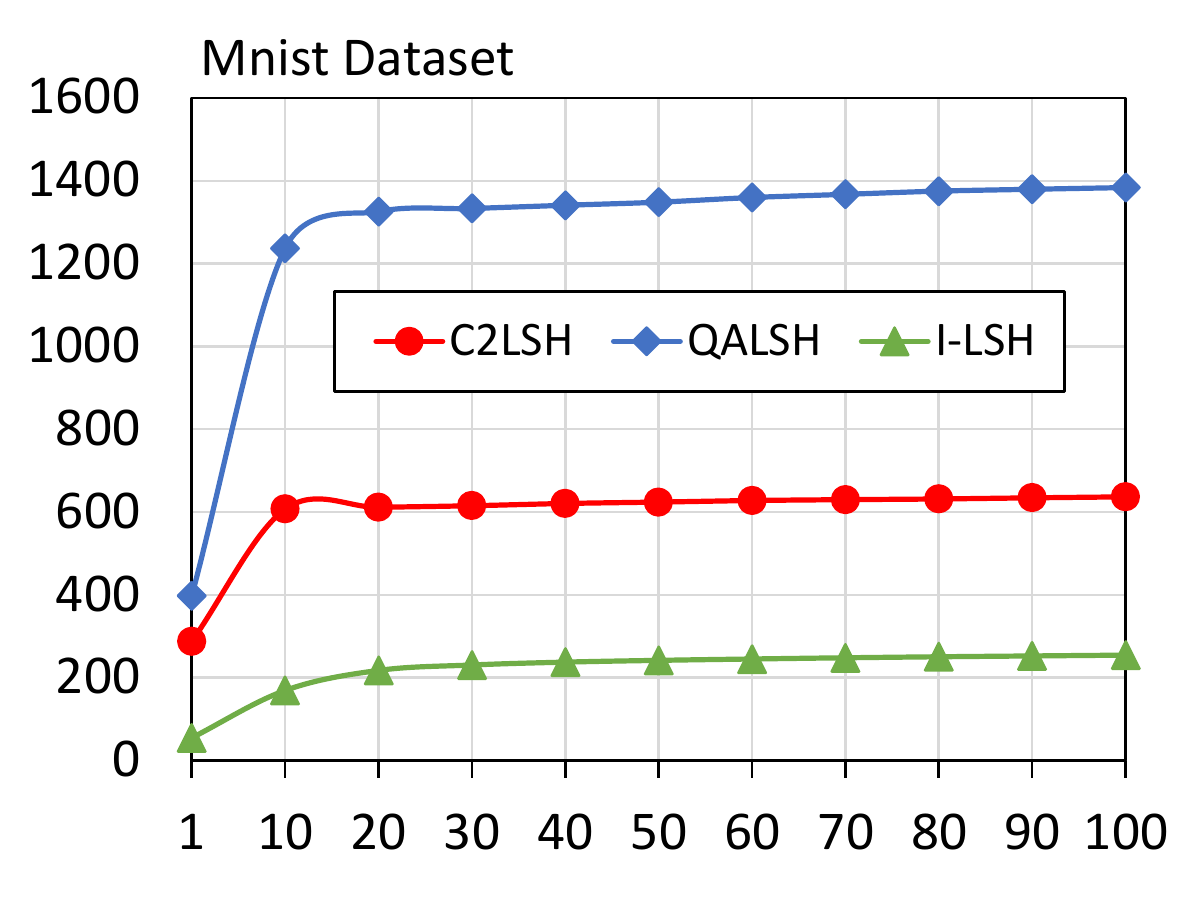}}
	\end{subfigure}\quad
	\begin{subfigure}[b]{0.31\textwidth}
		\centering
		{\includegraphics[width=\linewidth]{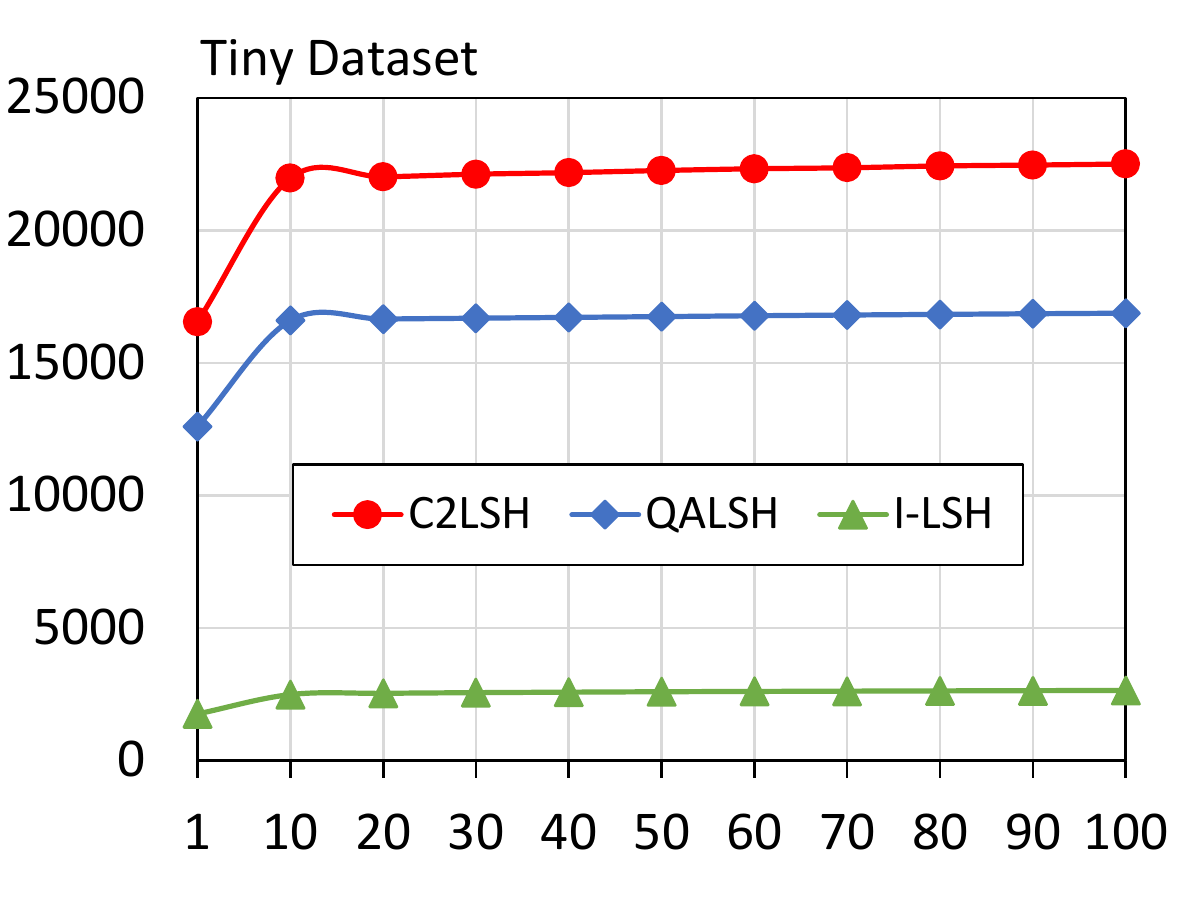}}
	\end{subfigure}
	
	\caption{Amount of Data Read (in MB) (Y axis) for $k$ (X Axis) on 6 datasets}
	\label{fig:expdiskIO}
\end{figure*}

\begin{figure*}[!h]
	\centering
	\begin{subfigure}[b]{0.31\textwidth}
		\centering
		{\includegraphics[width=\linewidth]{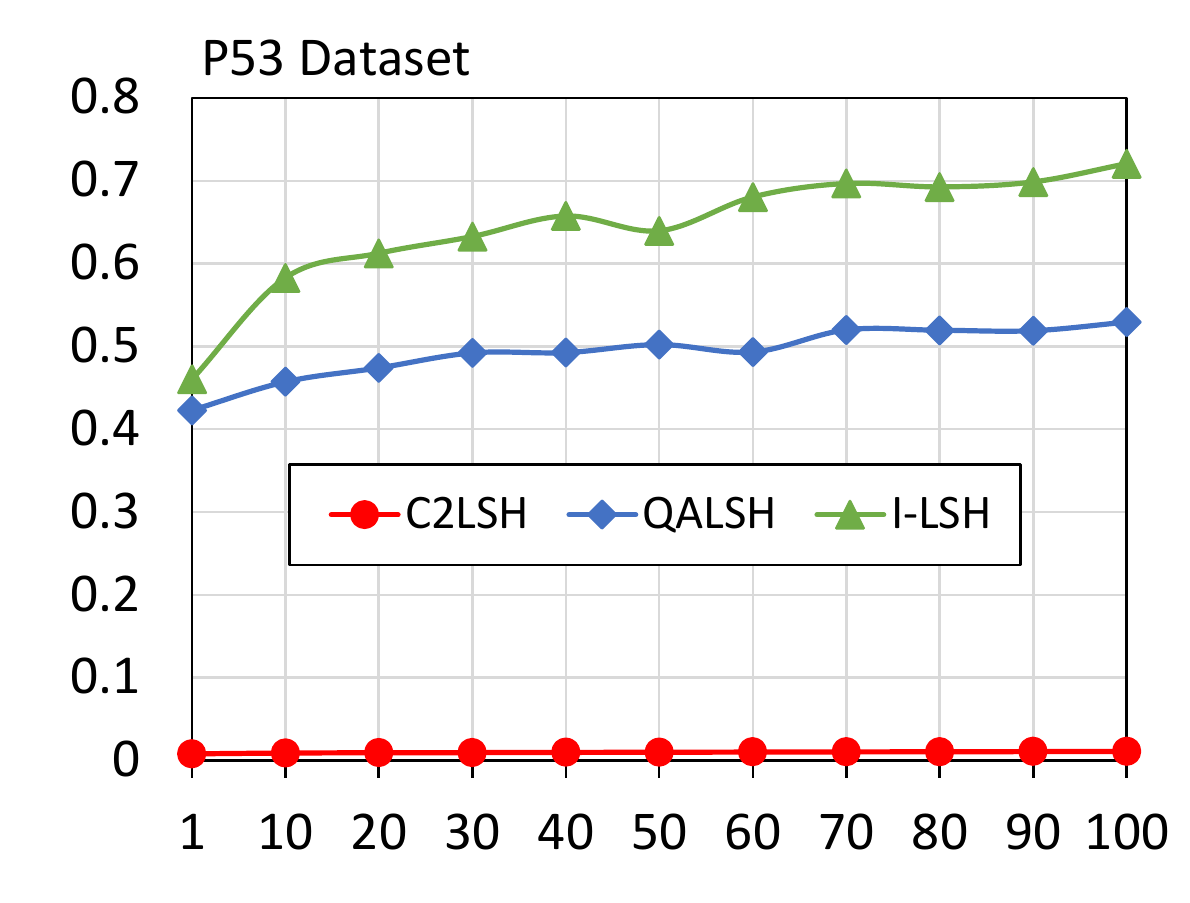}}
	\end{subfigure}\quad
	\begin{subfigure}[b]{0.31\textwidth}
		\centering
		{\includegraphics[width=\linewidth]{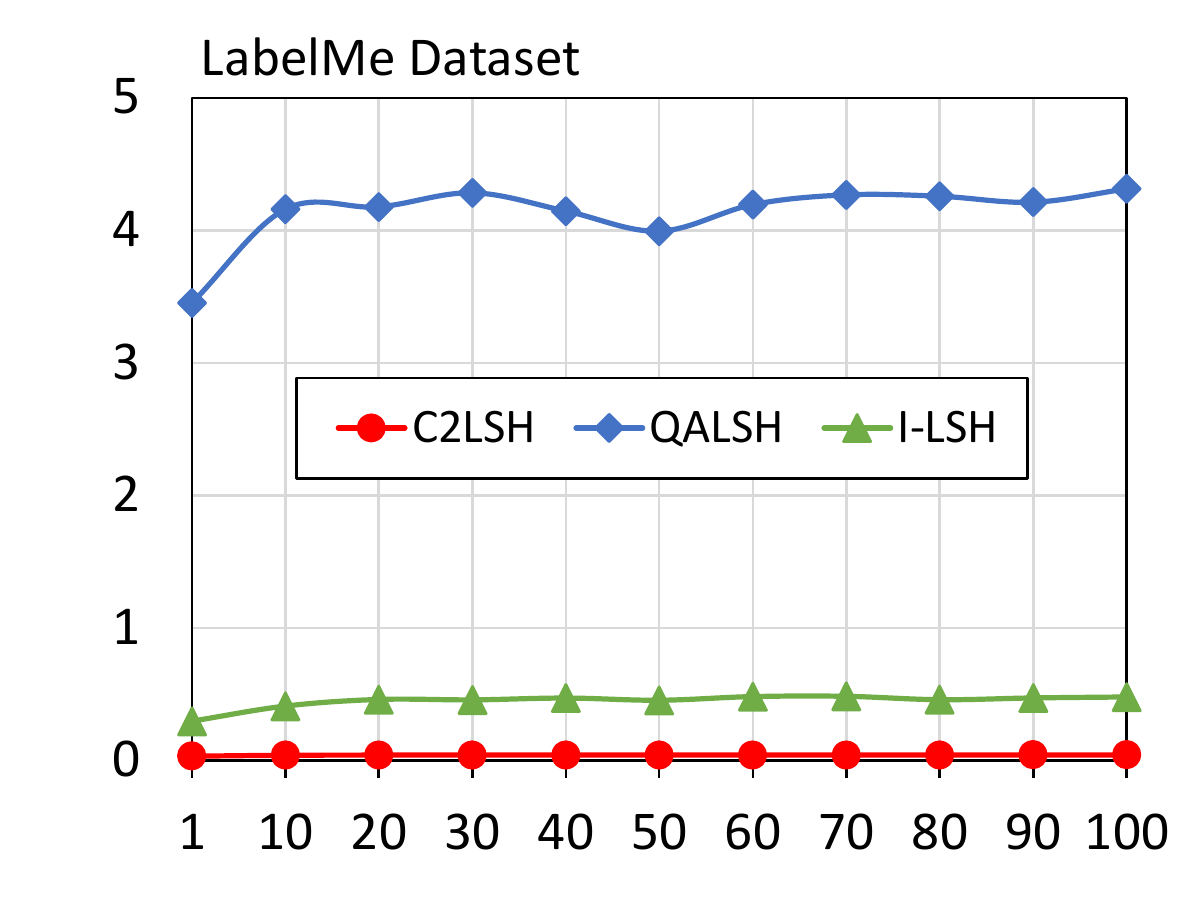}}
	\end{subfigure}\quad
	\begin{subfigure}[b]{0.31\textwidth}
		\centering
		{\includegraphics[width=\linewidth]{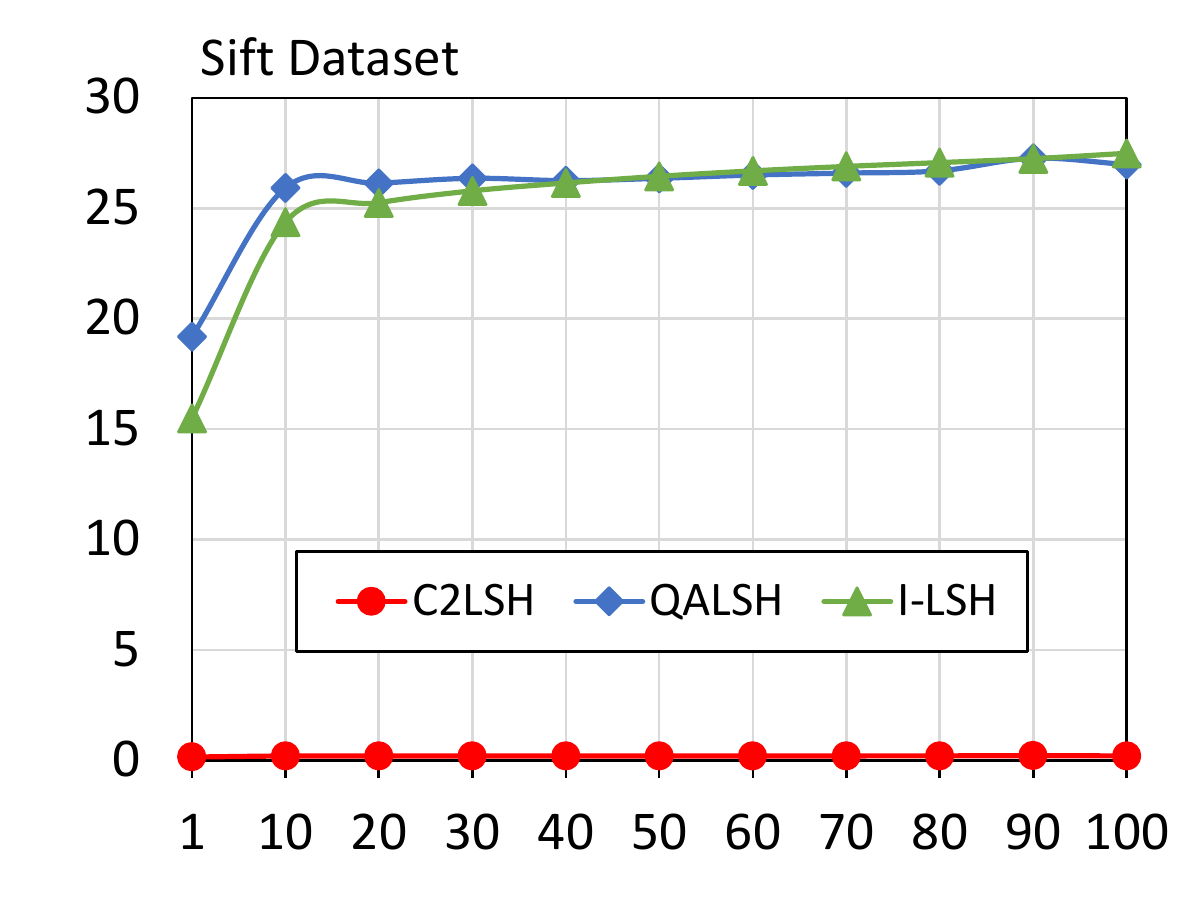}}
	\end{subfigure} \\
	
	\begin{subfigure}[b]{0.31\textwidth}
		\centering
		{\includegraphics[width=\linewidth]{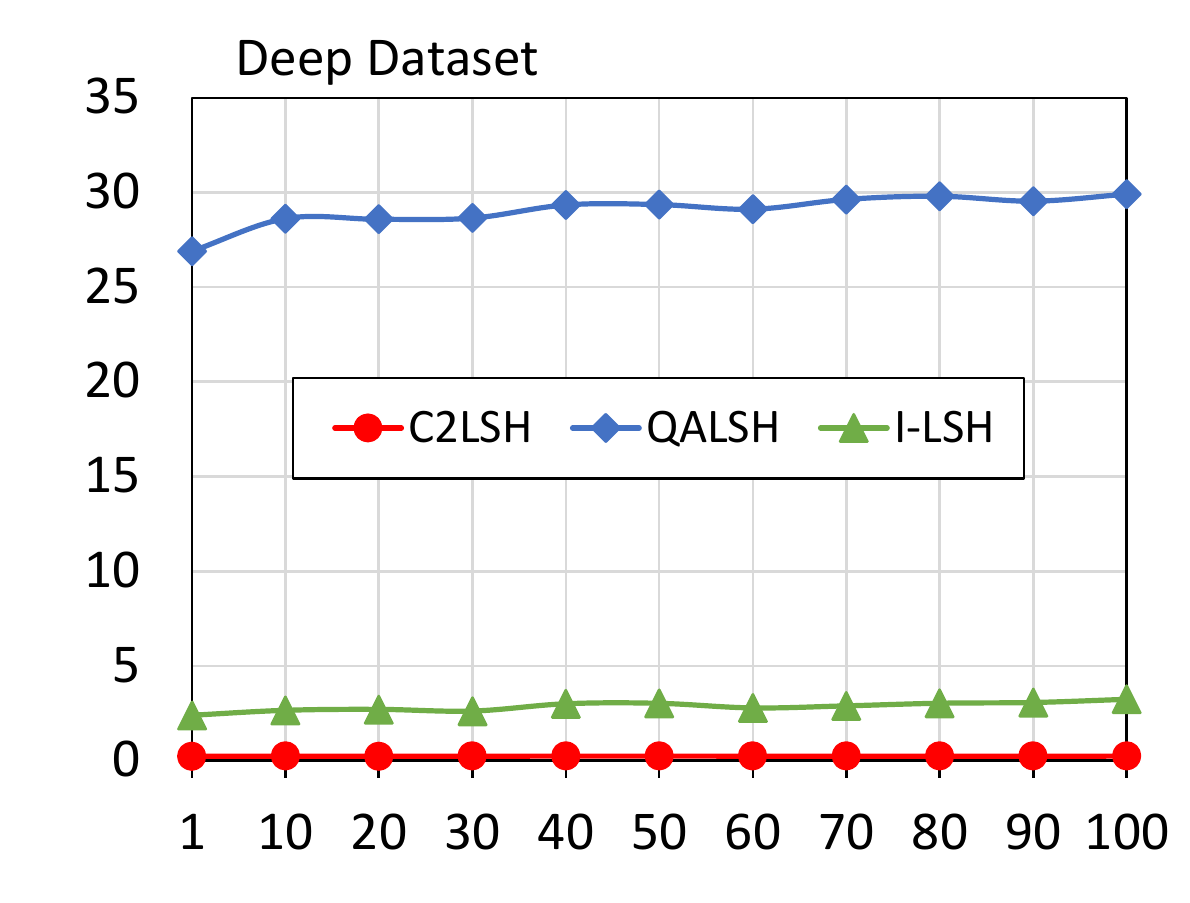}}
	\end{subfigure}\quad
	\begin{subfigure}[b]{0.31\textwidth}
		\centering
		{\includegraphics[width=\linewidth]{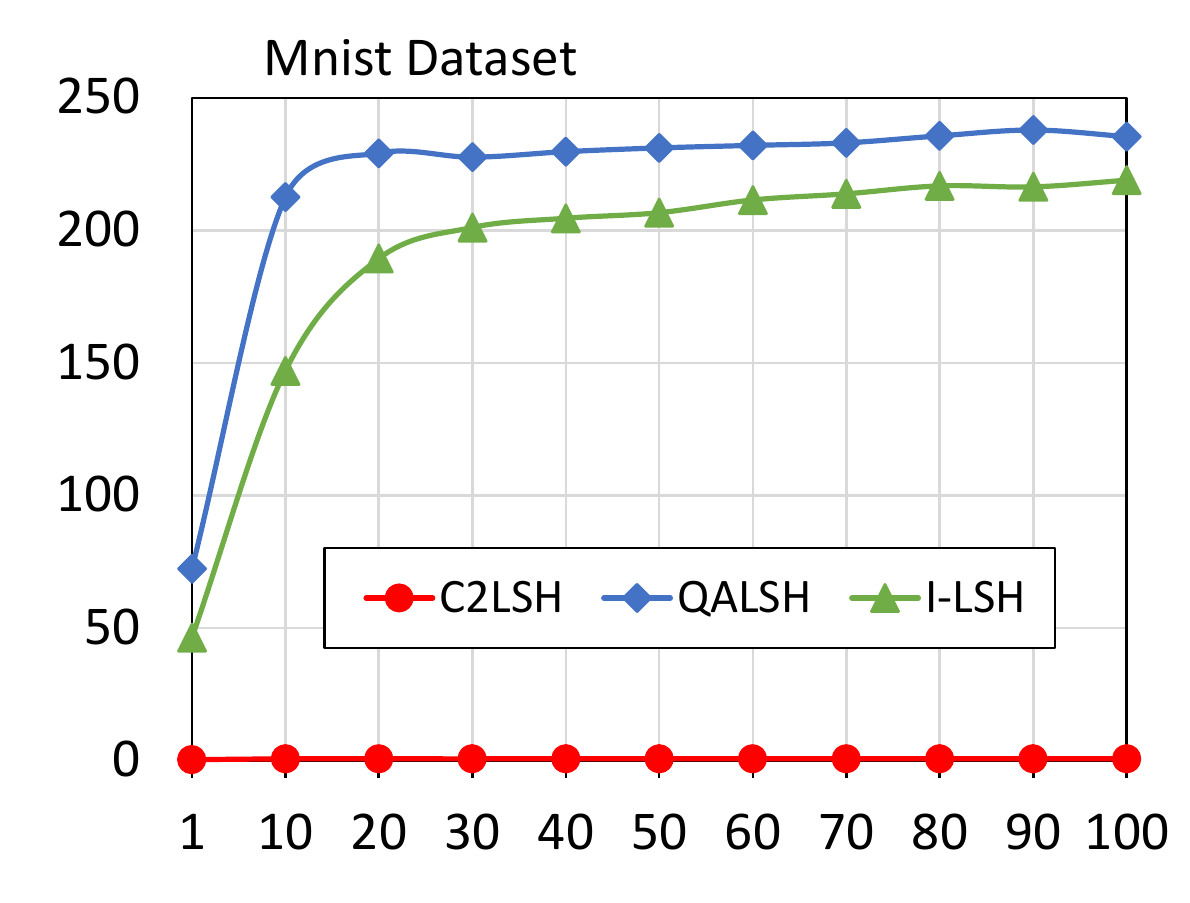}}
	\end{subfigure}\quad
	\begin{subfigure}[b]{0.31\textwidth}
		\centering
		{\includegraphics[width=\linewidth]{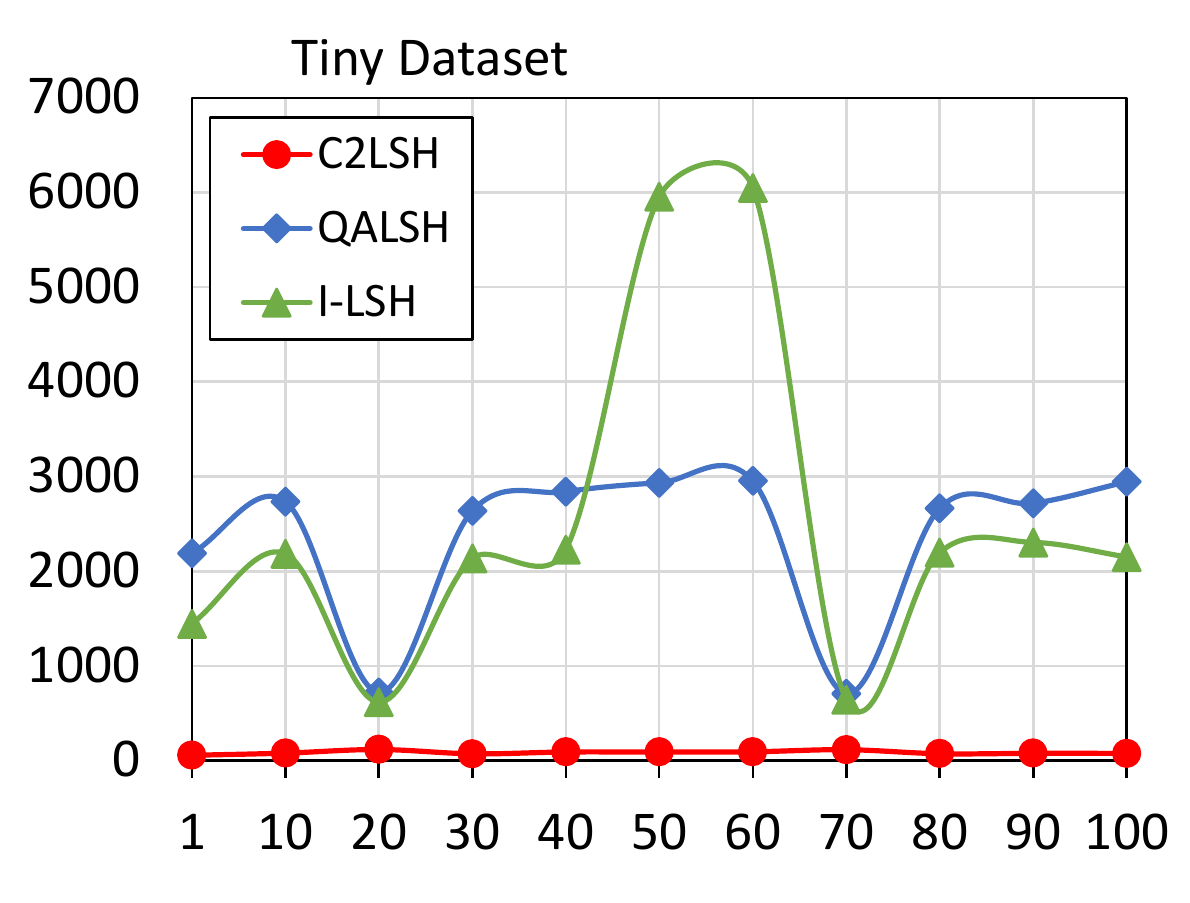}}
	\end{subfigure}
	
	\caption{Algorithm Time (in s) (Y axis) for $k$ (X Axis) on 6 datasets}
	\label{fig:expAlgTime}
\end{figure*}

\begin{figure*}[!h]
	\centering
	\begin{subfigure}[b]{0.31\textwidth}
		\centering
		{\includegraphics[width=\linewidth]{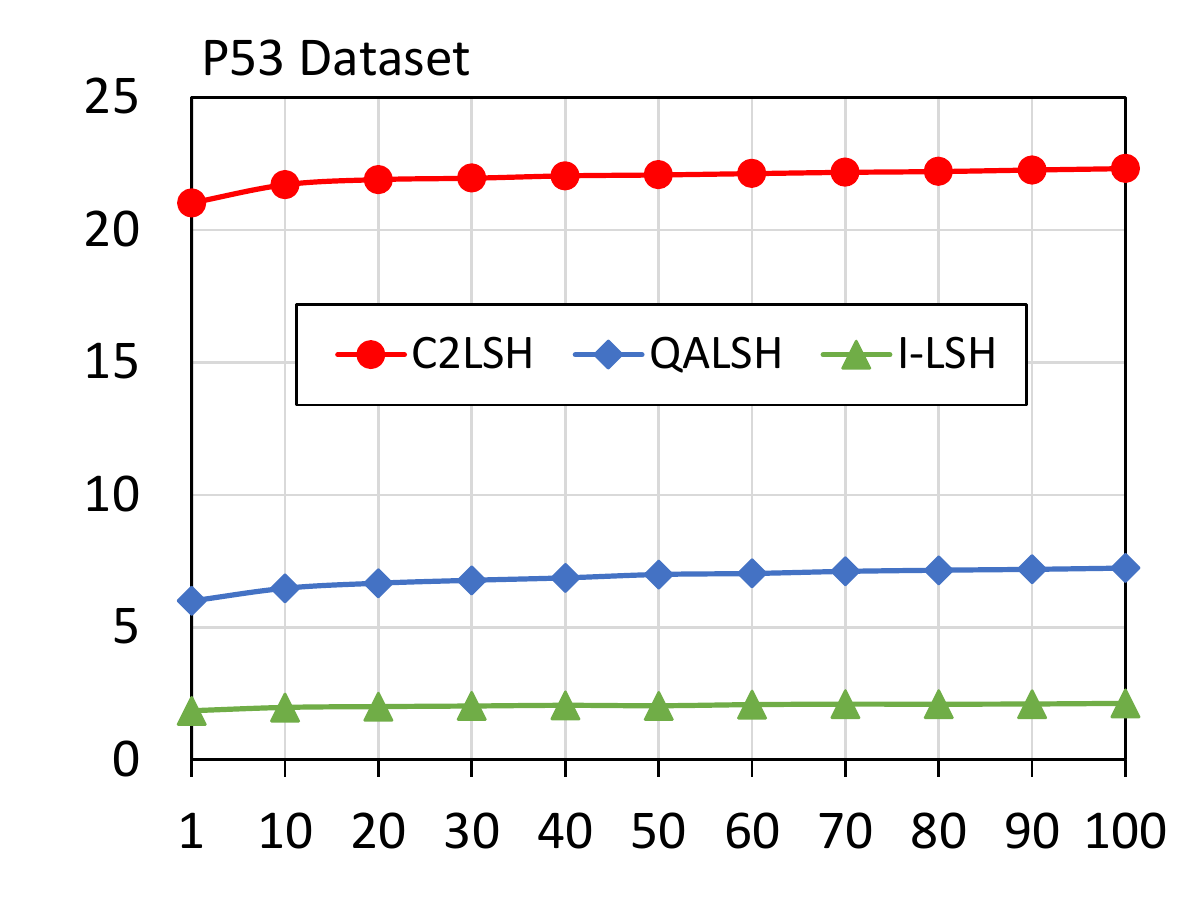}}
	\end{subfigure}\quad
	\begin{subfigure}[b]{0.31\textwidth}
		\centering
		{\includegraphics[width=\linewidth]{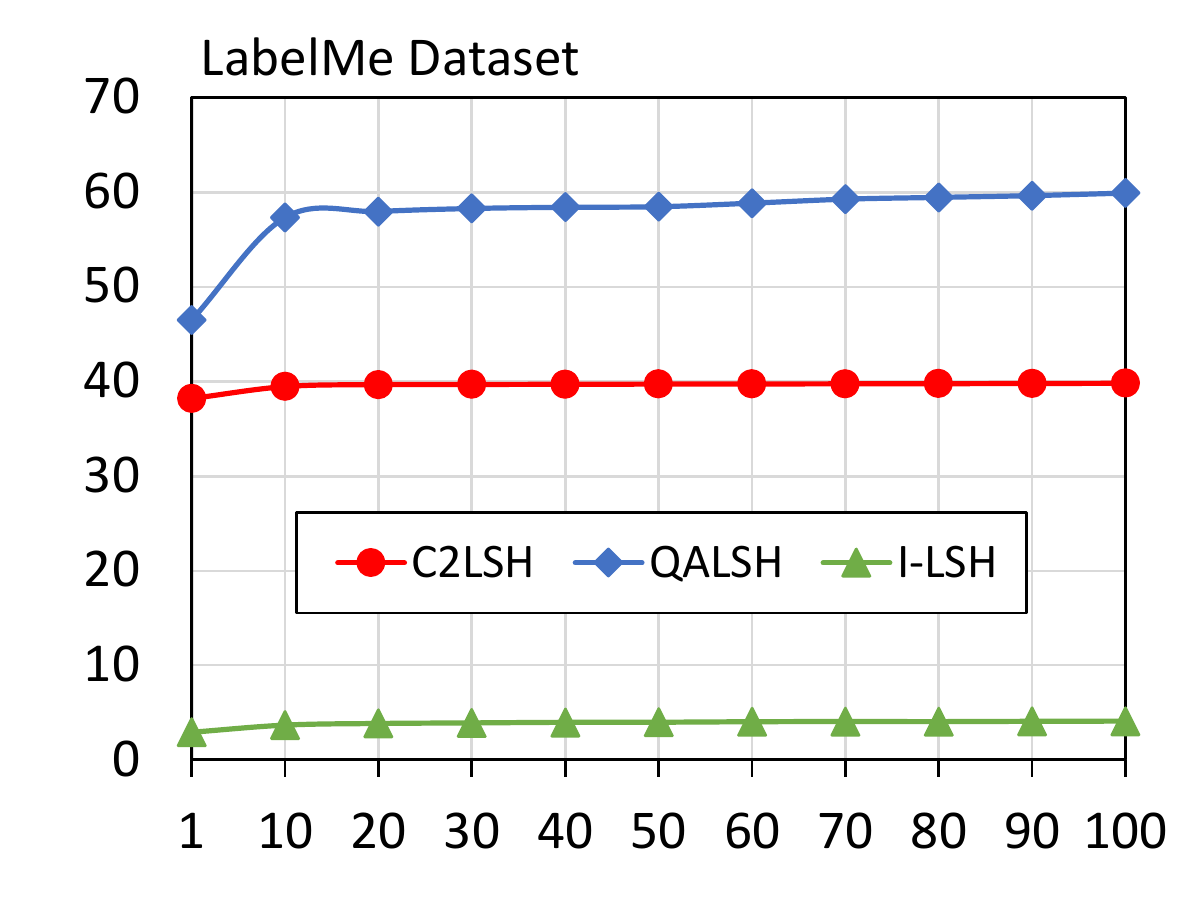}}
	\end{subfigure}\quad
	\begin{subfigure}[b]{0.31\textwidth}
		\centering
		{\includegraphics[width=\linewidth]{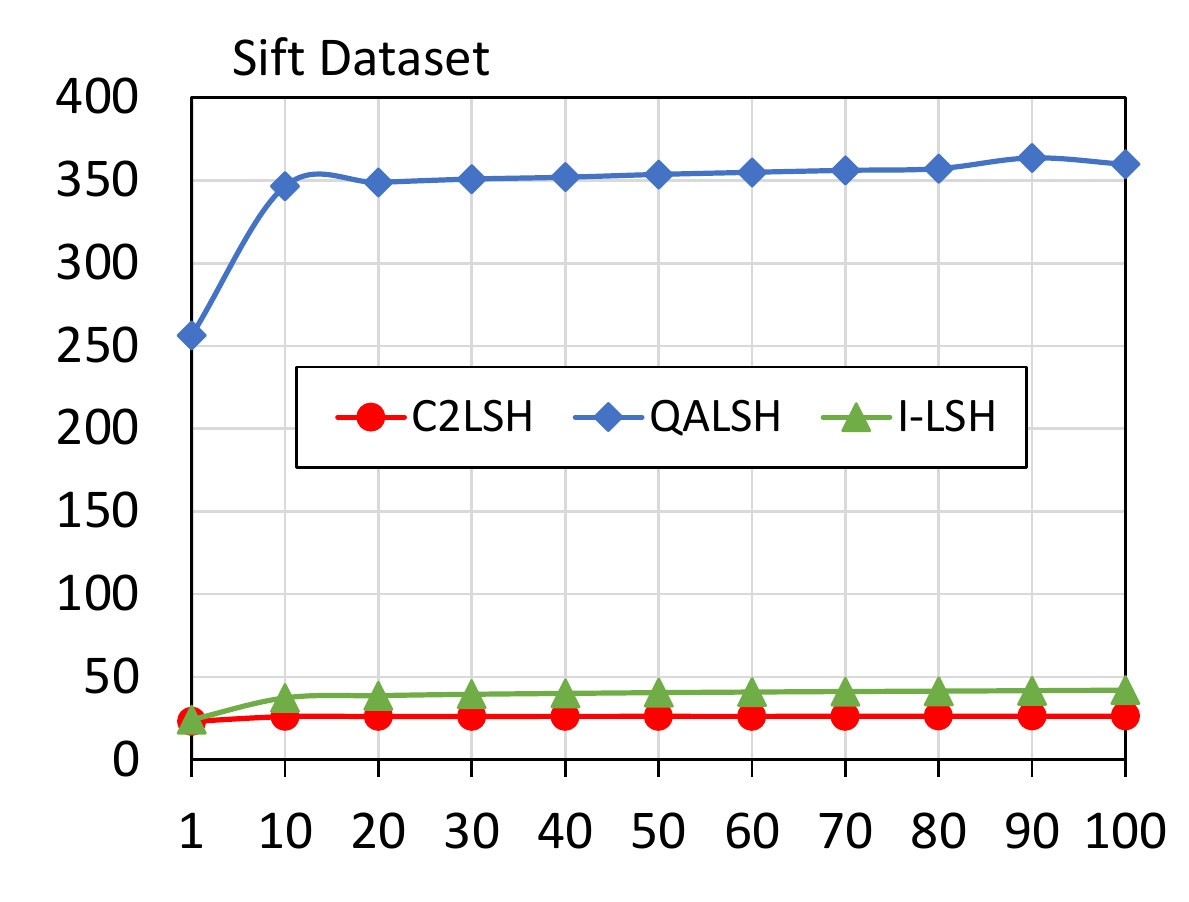}}
	\end{subfigure} \\
	
	\begin{subfigure}[b]{0.31\textwidth}
		\centering
		{\includegraphics[width=\linewidth]{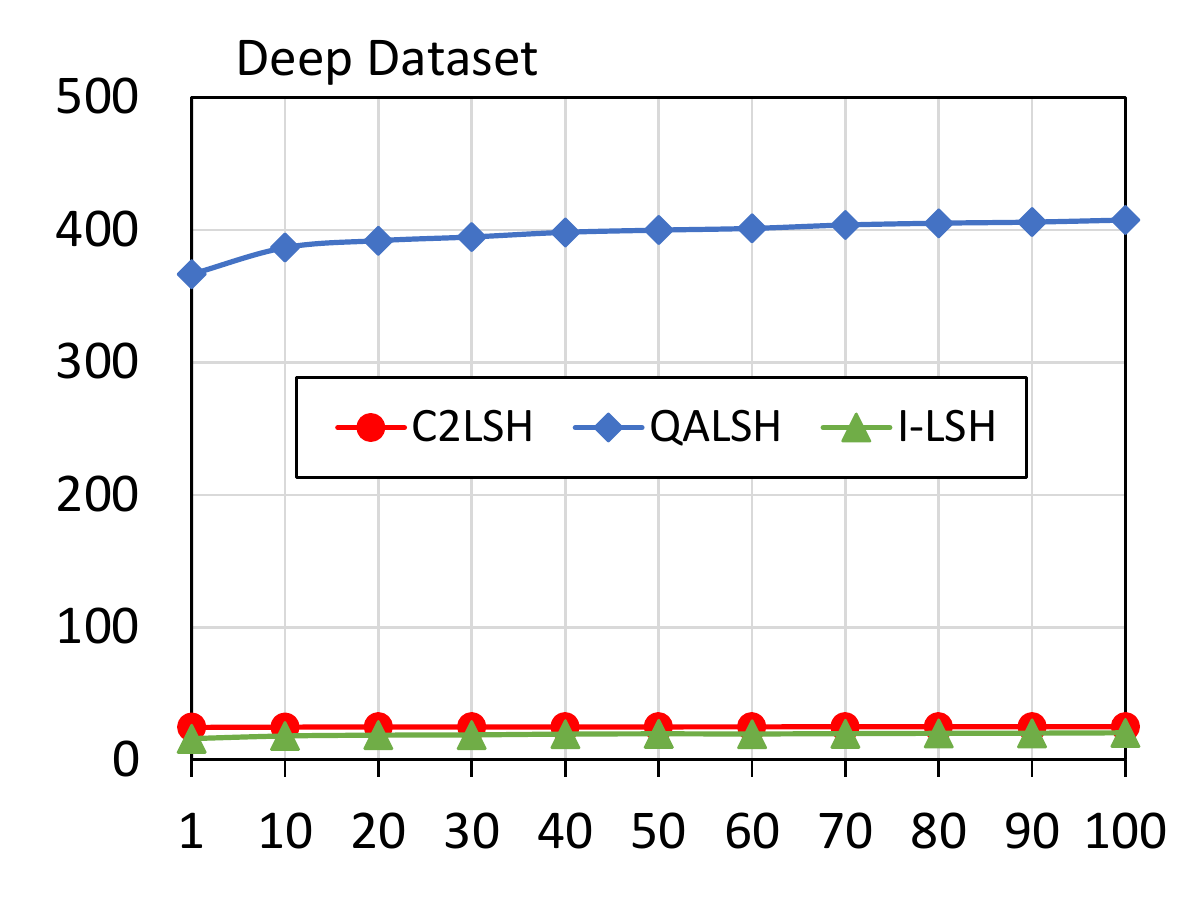}}
	\end{subfigure}\quad
	\begin{subfigure}[b]{0.31\textwidth}
		\centering
		{\includegraphics[width=\linewidth]{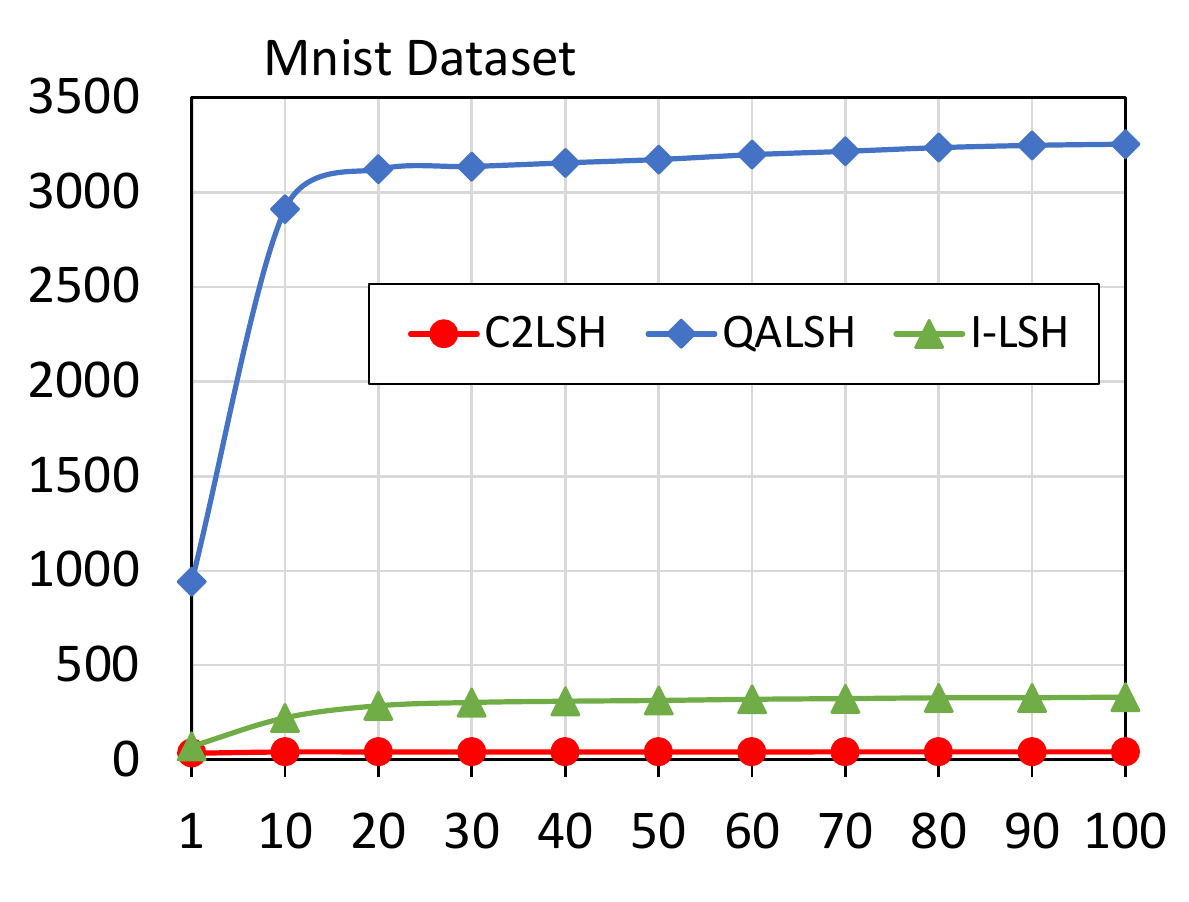}}
	\end{subfigure}\quad
	\begin{subfigure}[b]{0.31\textwidth}
		\centering
		{\includegraphics[width=\linewidth]{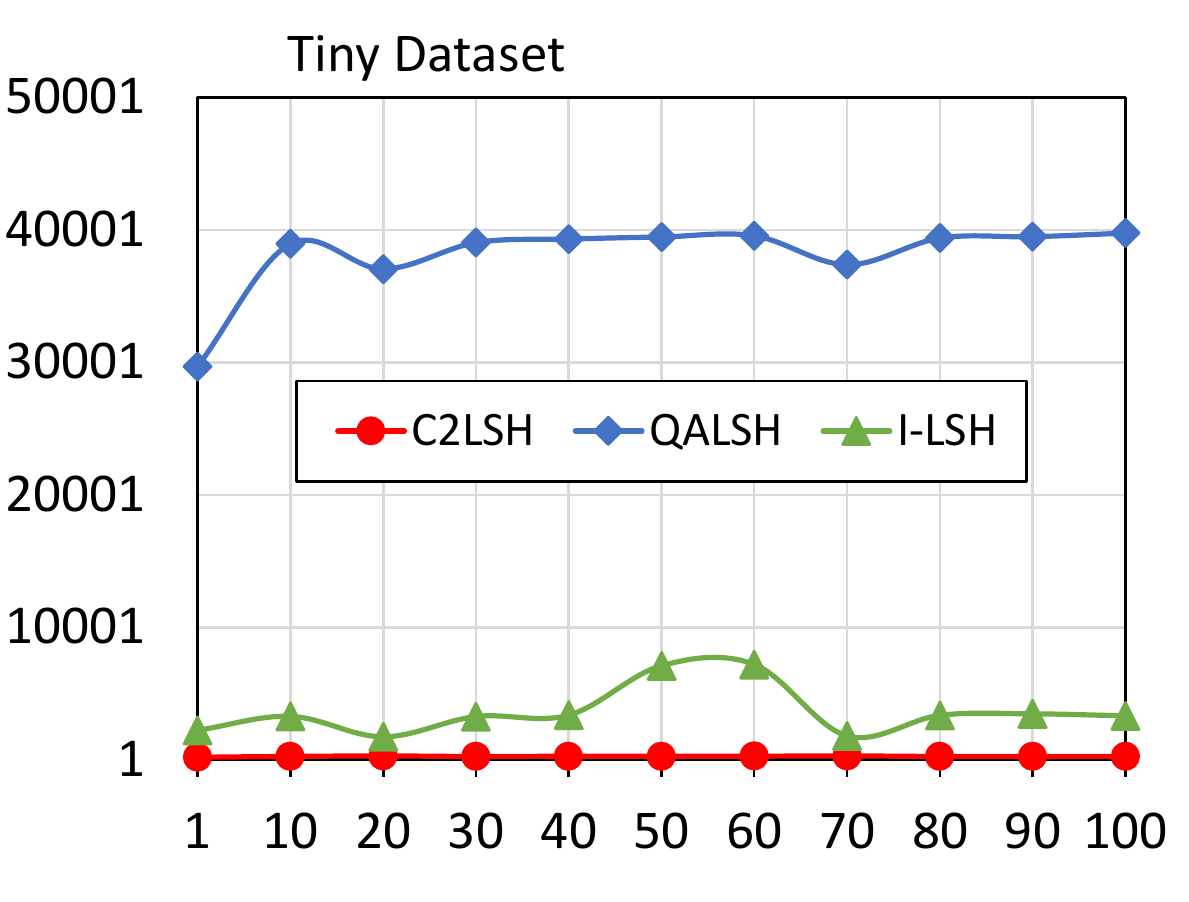}}
	\end{subfigure}
	
	\caption{HDD Query Processing Time (in s) (Y axis) for $k$ (X Axis) on 6 datasets}
	\label{fig:expTimeHDD}
\end{figure*}

\begin{figure*}[!h]
	\centering
	\begin{subfigure}[b]{0.31\textwidth}
		\centering
		{\includegraphics[width=\linewidth]{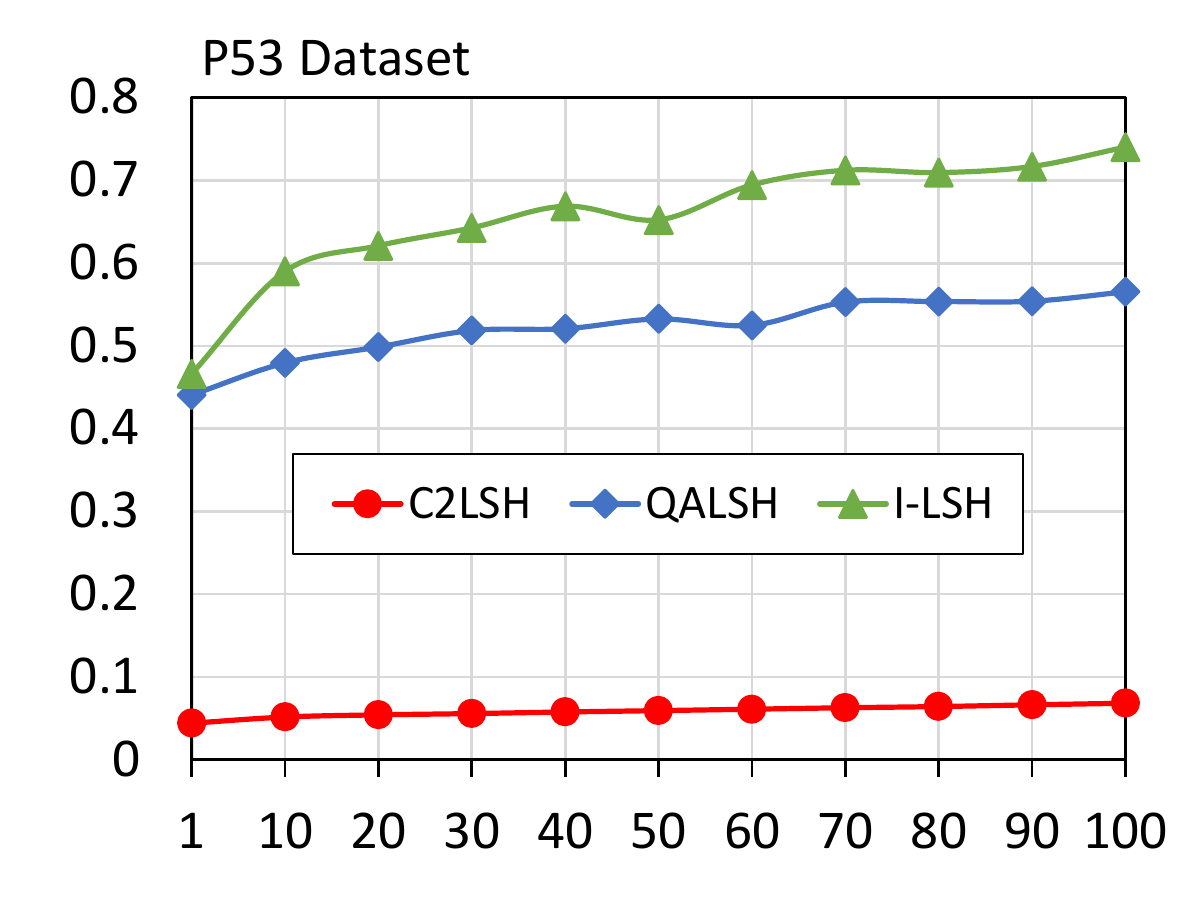}}
	\end{subfigure}\quad
	\begin{subfigure}[b]{0.31\textwidth}
		\centering
		{\includegraphics[width=\linewidth]{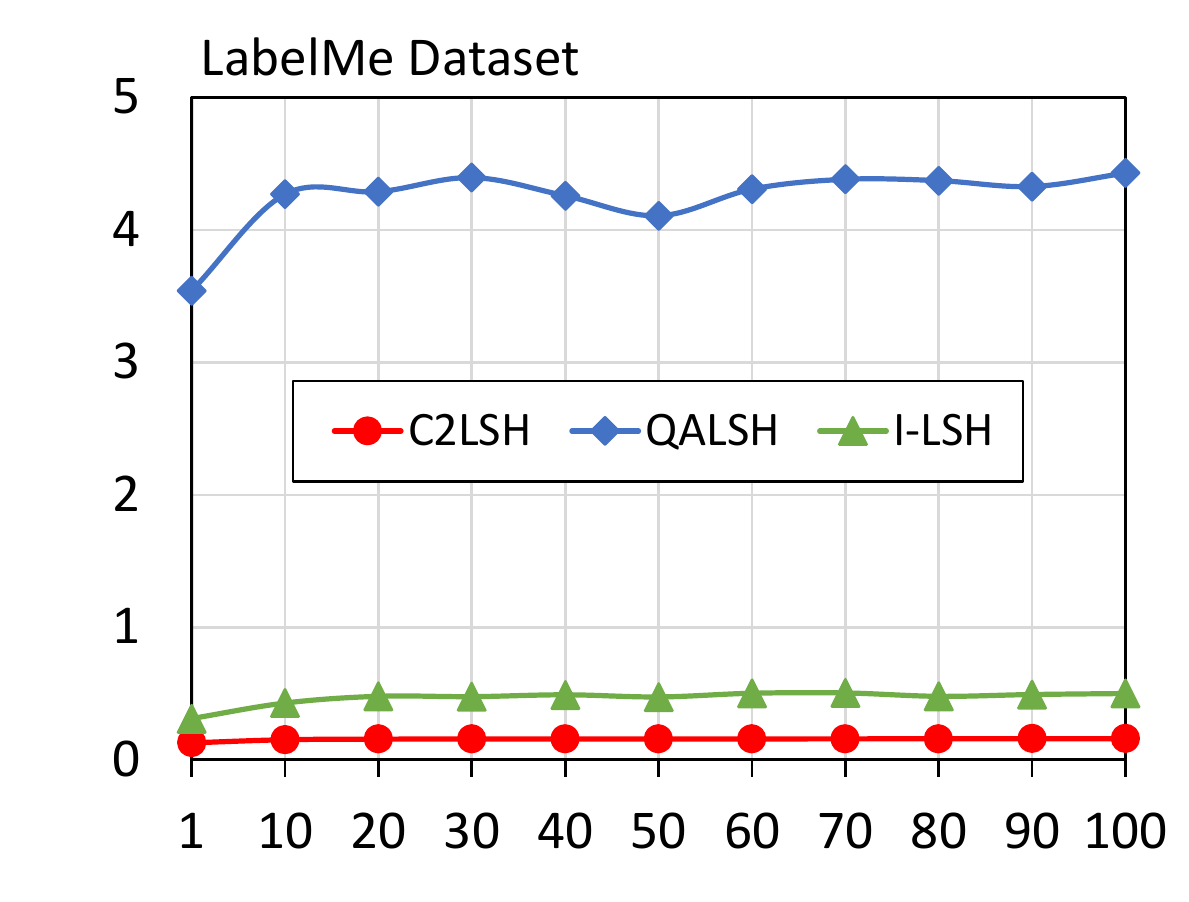}}
	\end{subfigure}\quad
	\begin{subfigure}[b]{0.31\textwidth}
		\centering
		{\includegraphics[width=\linewidth]{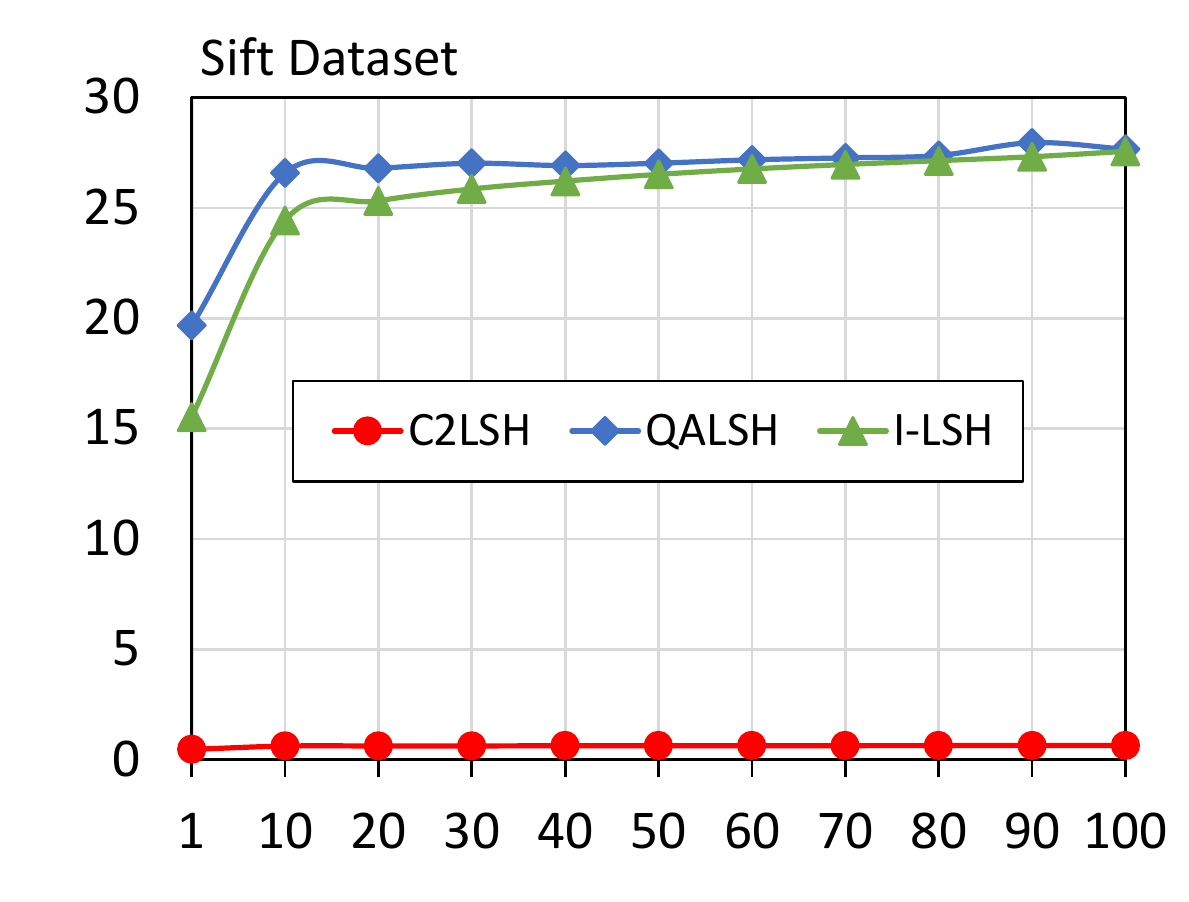}}
	\end{subfigure} \\
	
	\begin{subfigure}[b]{0.31\textwidth}
		\centering
		{\includegraphics[width=\linewidth]{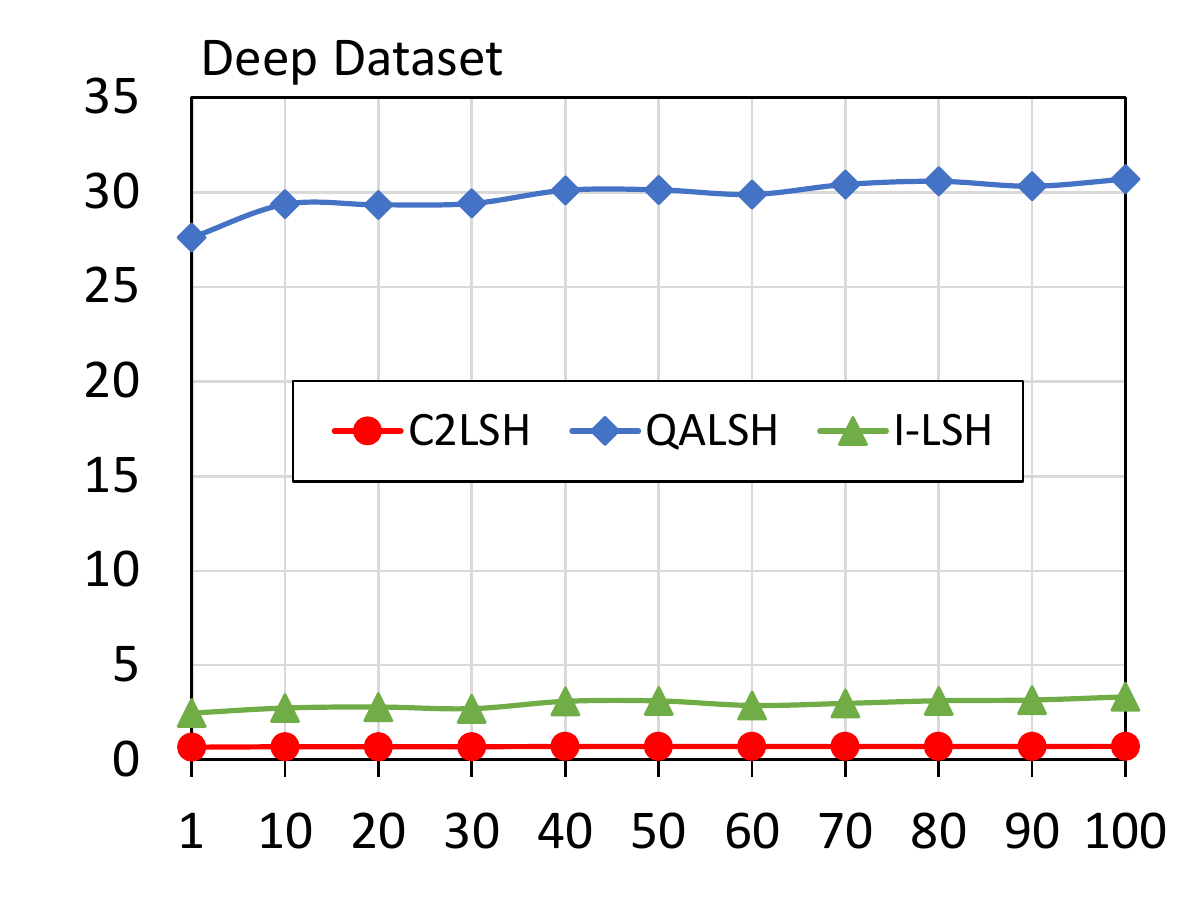}}
	\end{subfigure}\quad
	\begin{subfigure}[b]{0.31\textwidth}
		\centering
		{\includegraphics[width=\linewidth]{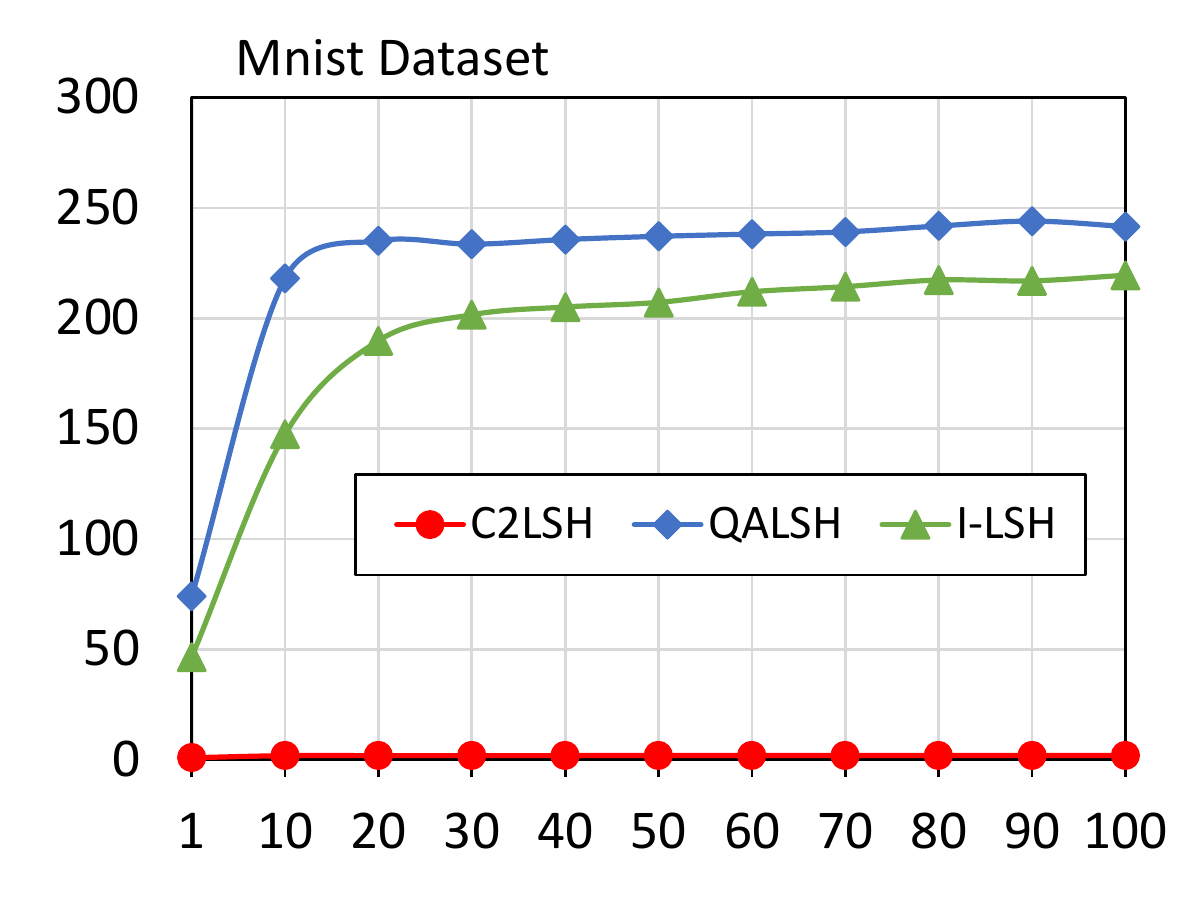}}
	\end{subfigure}\quad
	\begin{subfigure}[b]{0.31\textwidth}
		\centering
		{\includegraphics[width=\linewidth]{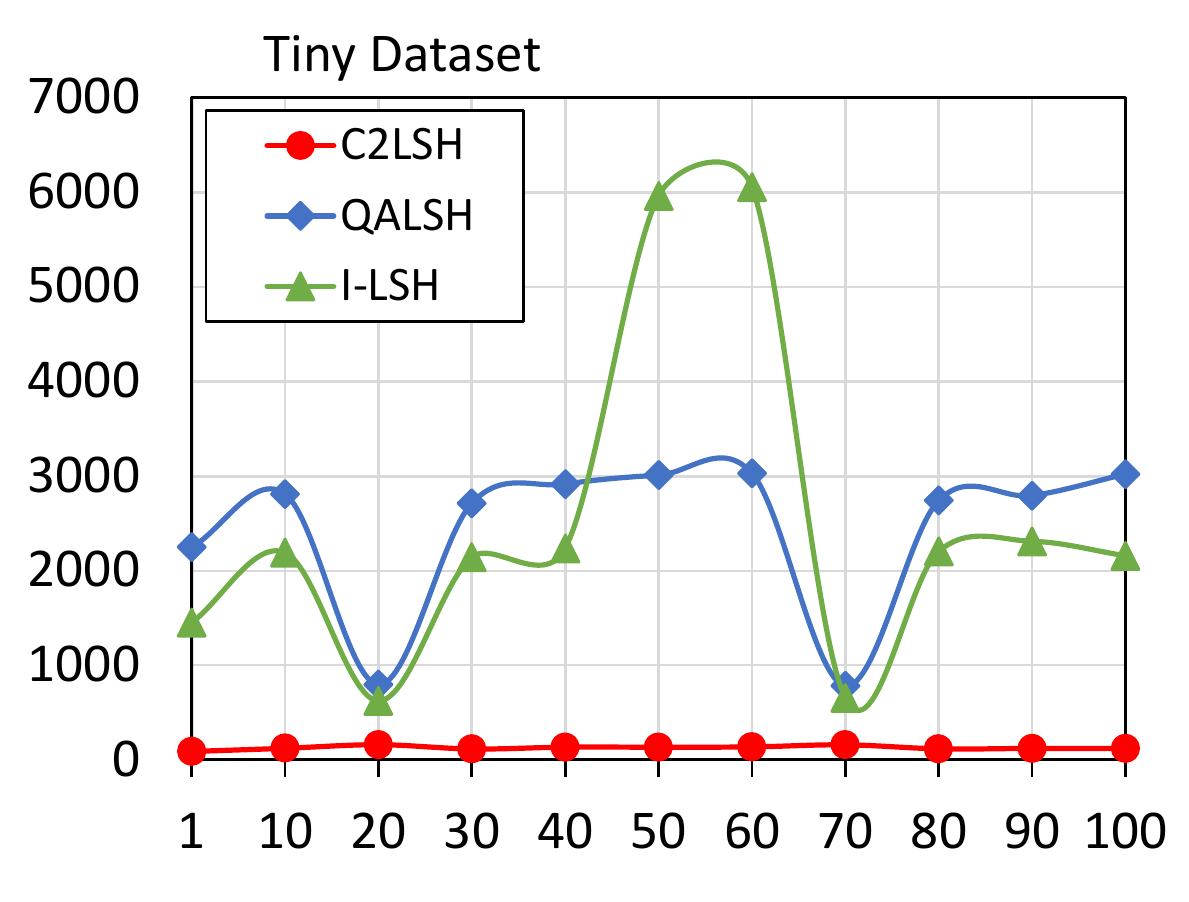}}
	\end{subfigure}
	
	\caption{SSD Query Processing Time (in s) (Y axis) for $k$ (X Axis) on 6 datasets}
	\label{fig:expTimeSSD}
\end{figure*}

\begin{figure*}[!h]
	\centering
	\begin{subfigure}[b]{0.31\textwidth}
		\centering
		{\includegraphics[width=\linewidth]{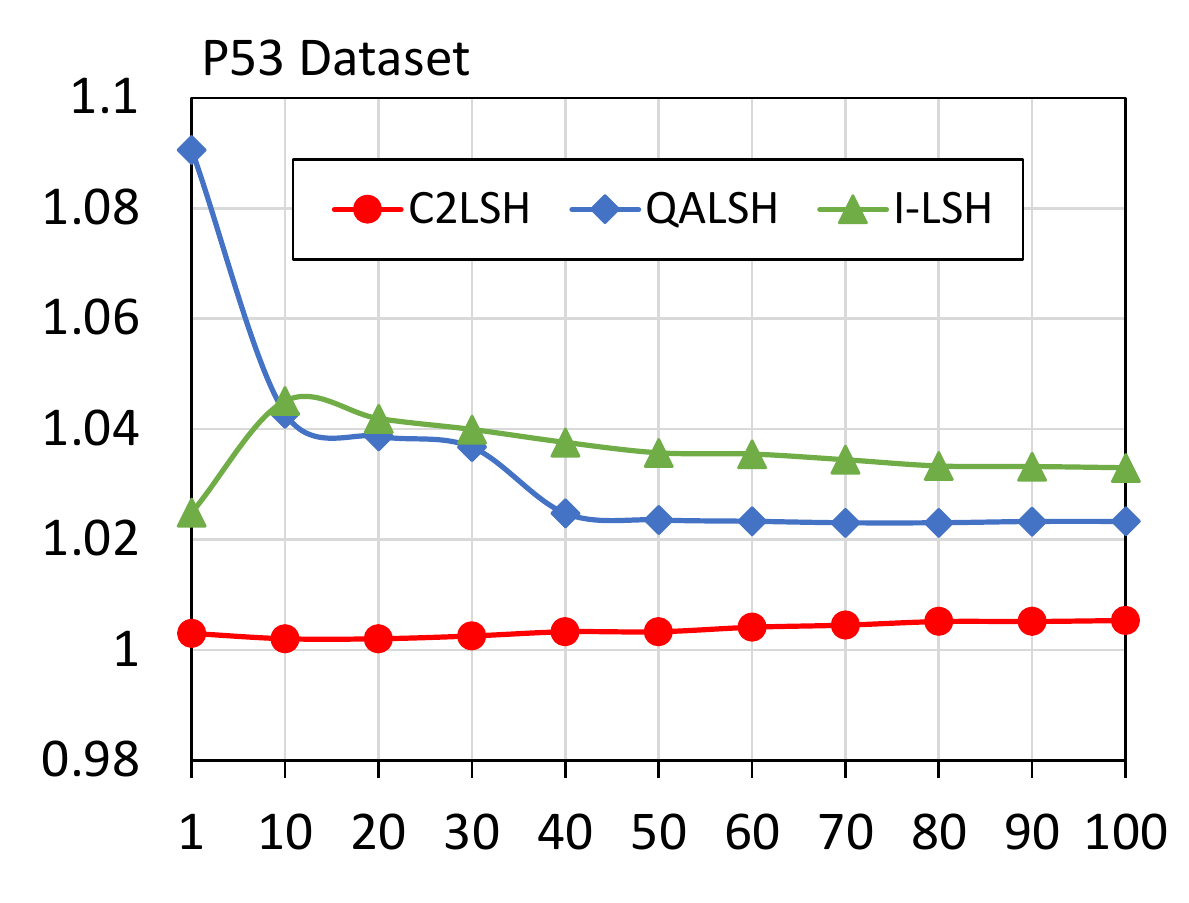}}
	\end{subfigure}\quad
	\begin{subfigure}[b]{0.31\textwidth}
		\centering
		{\includegraphics[width=\linewidth]{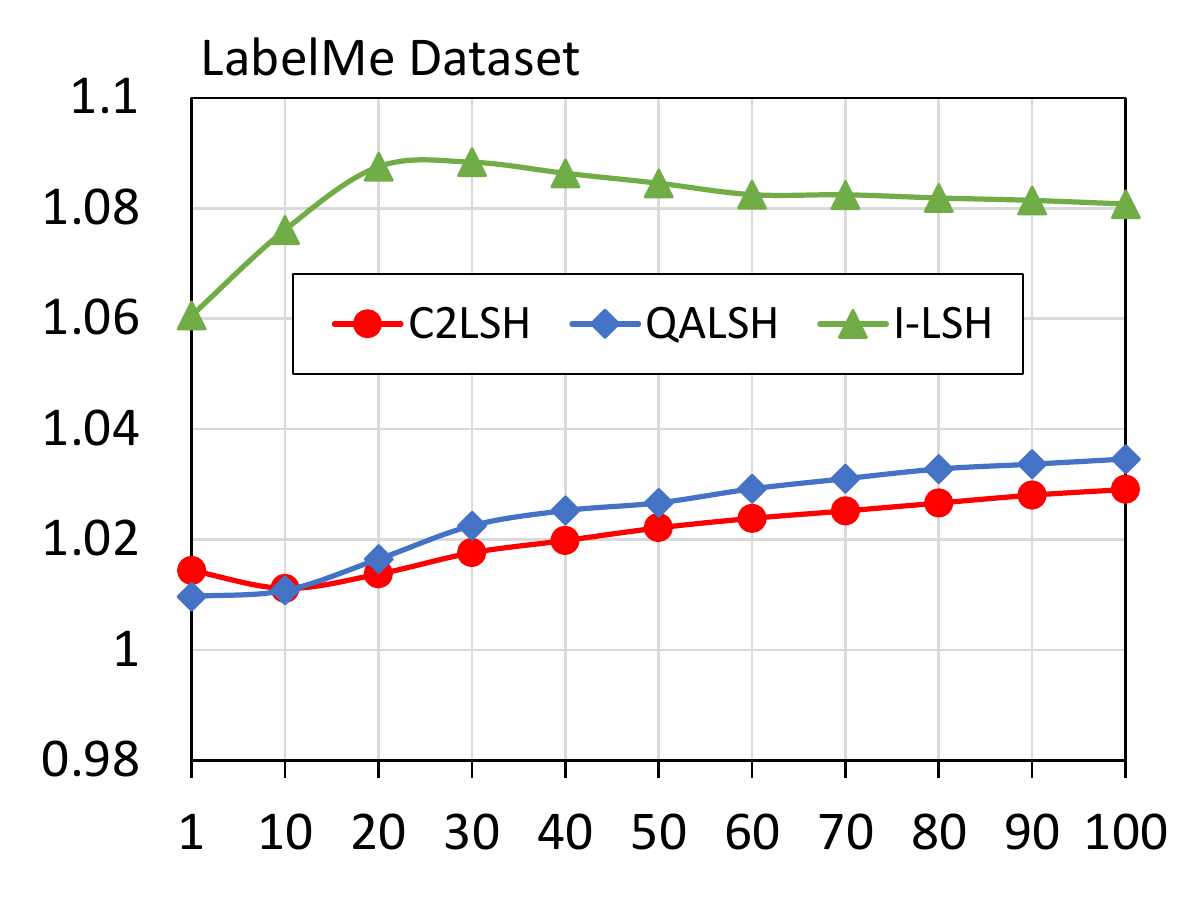}}
	\end{subfigure}\quad
	\begin{subfigure}[b]{0.31\textwidth}
		\centering
		{\includegraphics[width=\linewidth]{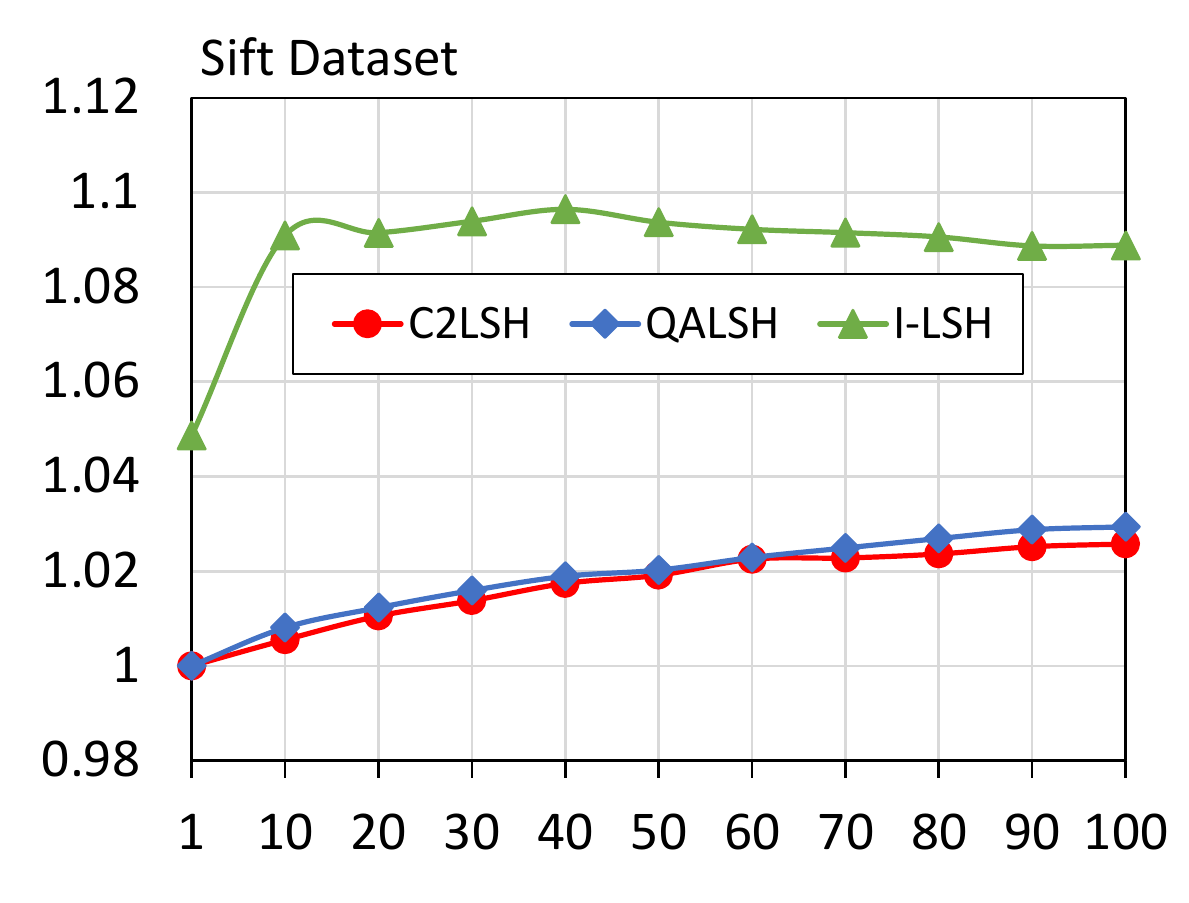}}
	\end{subfigure} \\
	
	\begin{subfigure}[b]{0.31\textwidth}
		\centering
		{\includegraphics[width=\linewidth]{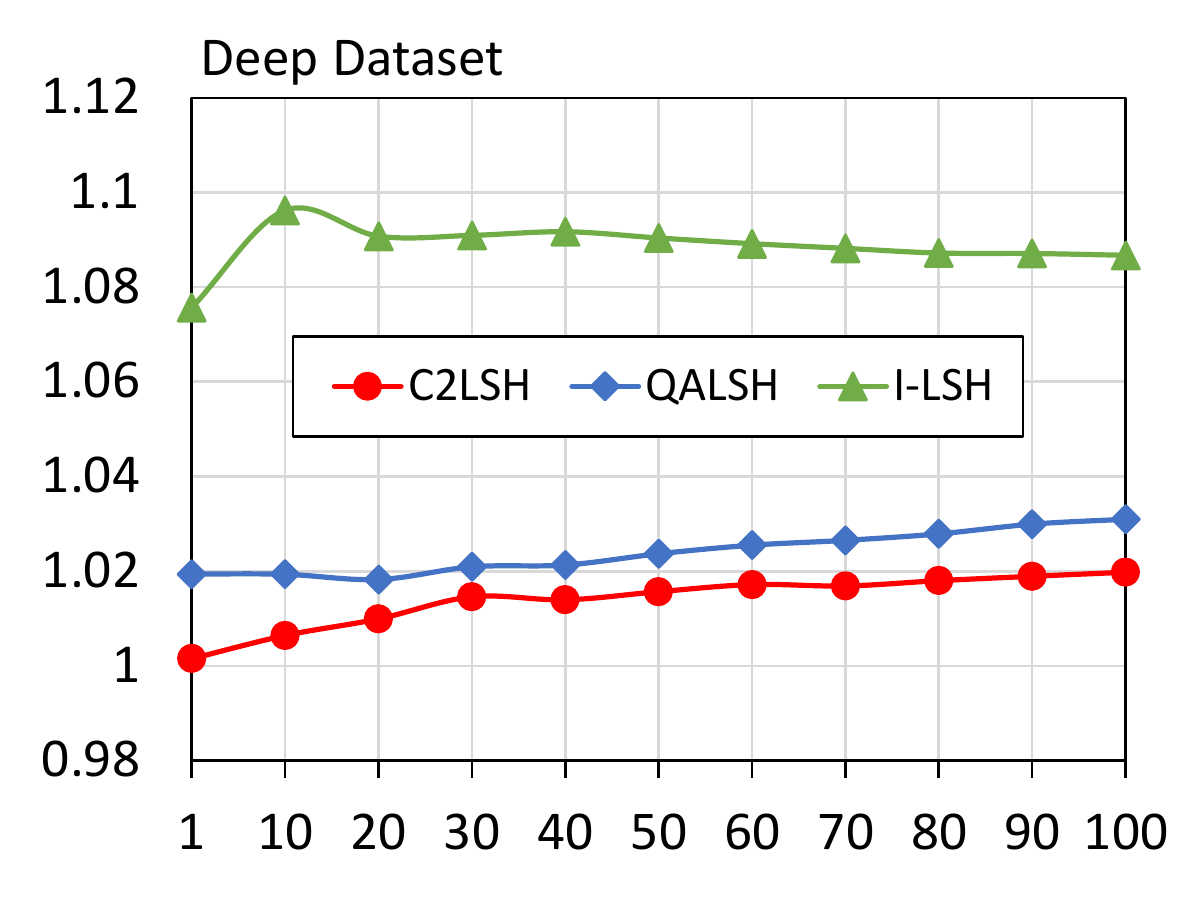}}
	\end{subfigure}\quad
	\begin{subfigure}[b]{0.31\textwidth}
		\centering
		{\includegraphics[width=\linewidth]{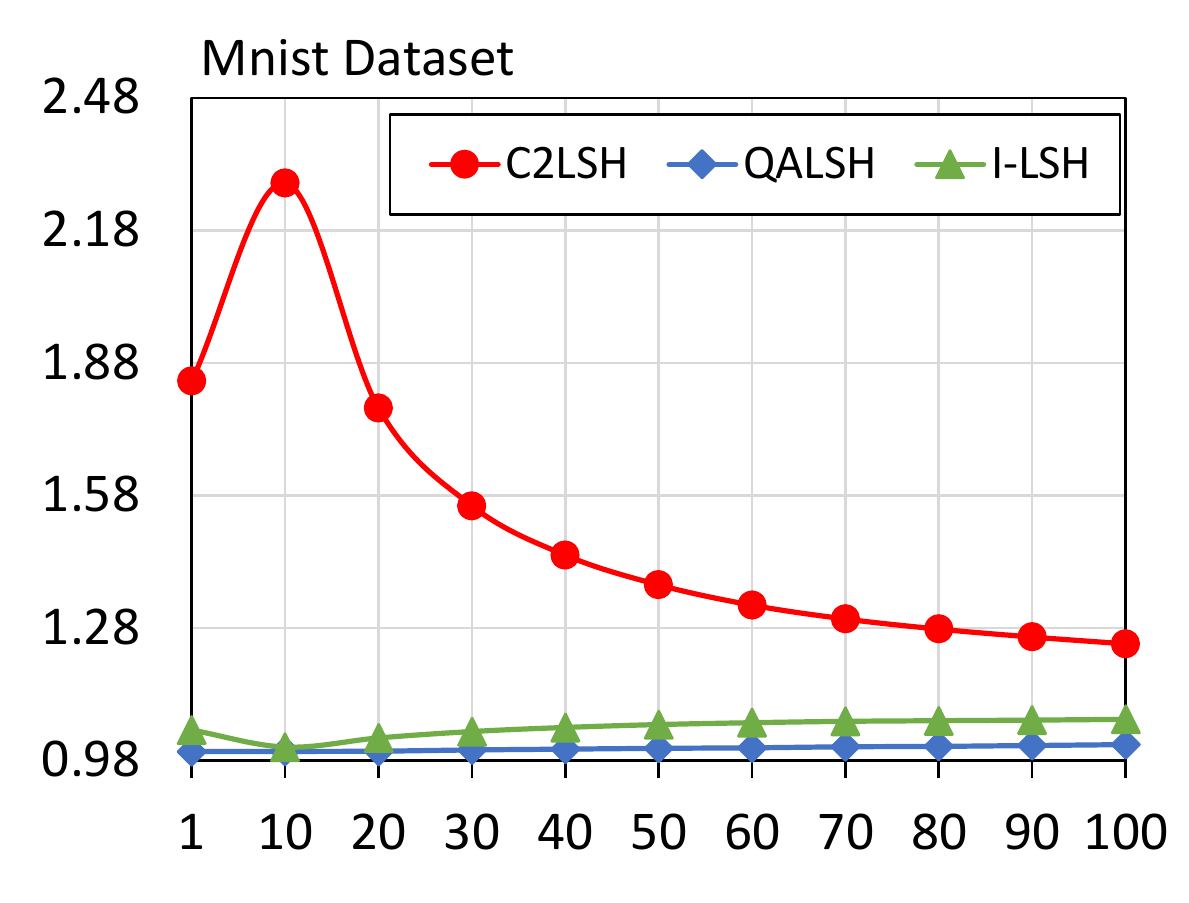}}
	\end{subfigure}\quad
	\begin{subfigure}[b]{0.31\textwidth}
		\centering
		{\includegraphics[width=\linewidth]{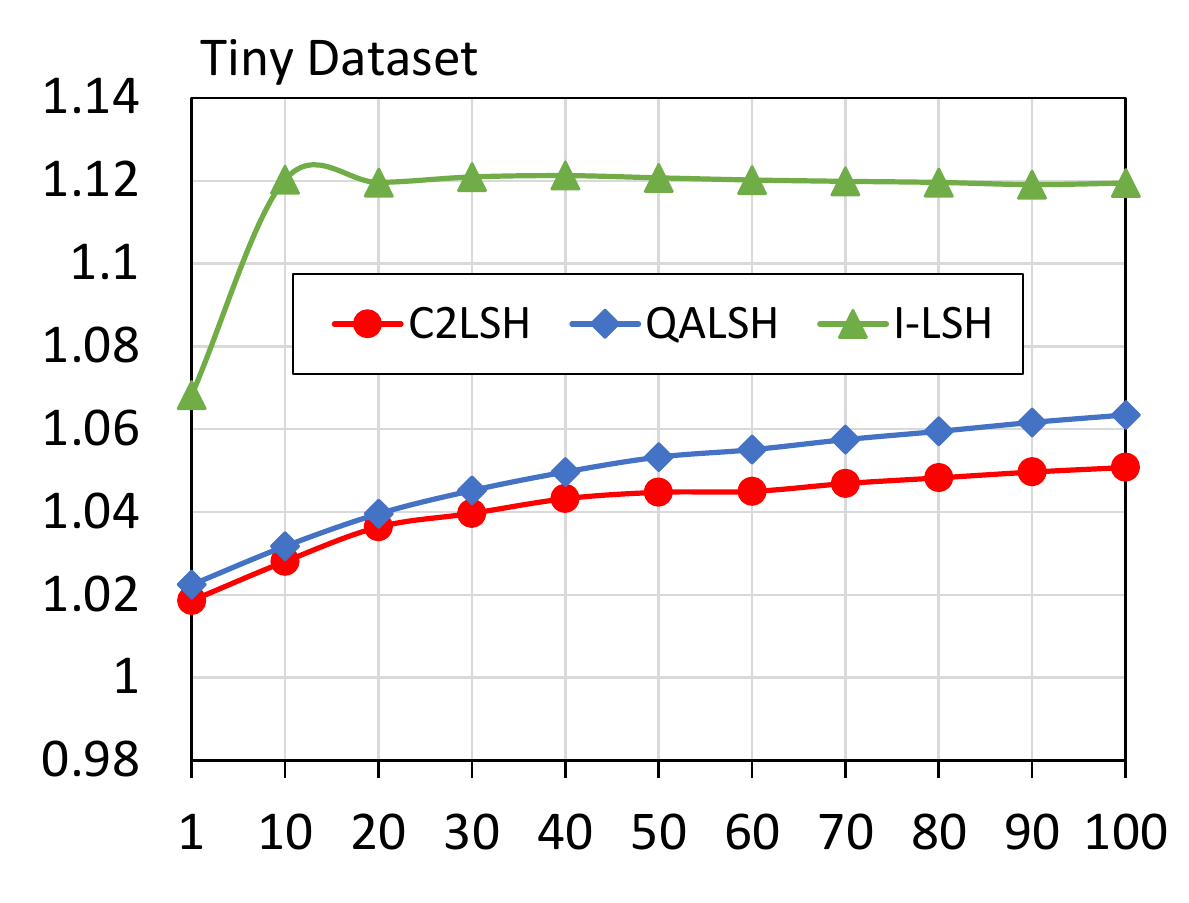}}
	\end{subfigure}
	
	\caption{Accuracy Ratio (Y axis) for different $k$ (X Axis) on 6 datasets}
	\label{fig:expRatio}
\end{figure*}

\subsection{Discussion of the Performance Results}

\noindent\textbf{Number of Disk Seeks:} Figure \ref{fig:expdiskseek} shows the required number of disk seeks (random I/Os) by the experimented techniques. The interesting observation is that I-LSH performs the best for P53, LabelMe, Sift, and Deep datasets. However, its performance degrades as the dataset size becomes large (i.e. greater than approximately one million points). This is because I-LSH needs to find the closest projected point each time the radius needs to be expanded, which further requires reading the indexed points from the disk several times. We also observe that QALSH has a better performance compared to C2LSH for smaller datasets (i.e. P53), but as the dataset size (number of points) increases, the number of seeks are significantly higher than C2LSH and I-LSH. This is happening because the search radiuses of QALSH are larger than C2LSH in larger datasets, which results in more radius expansions, which further results in higher disk seeks. \\

\noindent\textbf{Amount of Data Read:} Figure \ref{fig:expdiskIO} shows the total amount of data that was read from the index files. I-LSH always has the least amount of data read for all datasets because it incrementally searches for the nearest points in the projections instead of having buckets and fixed widths. However, we later show that these I/O savings are offset by the processing time of finding these nearest points. C2LSH reads more data than QALSH for most datasets (except Mnist) because it has more projections to process (since QALSH uses less hash projections because they are query-aware). \\

\noindent\textbf{Algorithm Time:} Figure \ref{fig:expAlgTime} shows the time needed by an algorithm to find the candidates (excluding the I/O times). This figure shows the huge overhead of I-LSH which is caused due to their incremental searching for the nearest projected neighbors. Also, since I-LSH and QALSH both use B+-trees, which become huge for the larger datasets, their performance degrades heavily in these cases while searching for candidates. Since C2LSH does not have any overhead of additional index structures (such as B+-tree), it has the least Algorithm time for all datasets. In terms of Algorithm Time, I-LSH is faster than QALSH (except for the P53 dataset - which is the smallest dataset in our experiments) mainly because it has to process less hash functions than QALSH \cite{Liu:2019}. \\

\noindent\textbf{False Positive Removal Time:} We also analyzed the time it takes to read the actual data point from the external memory in order to calculate Euclidean distance with the query (for removing false positives). Since all three algorithms have an upper bound of the number of candidates ($k+100$) it produces, all algorithms took similar time which was less than 0.5 ms. Due to space limitations, we do not show these results. \\ 

\noindent\textbf{Query Processing Time (on HDD):} Figure \ref{fig:expTimeHDD} shows the overall time required to solve a given k-NN query on a Hard Disk Drive. I-LSH performs the best for smaller datasets (P53 and LabelMe) because its Algorithm Time overhead is small, but as the dataset size increases, the Algorithm Time overhead offsets the savings in disk seeks and performs worse than C2LSH (but better than QALSH). Except for the smallest dataset (P53), QALSH is the slowest of the three algorithms. It works good for smaller datasets (P53) but does not scale well for moderate and large sized datasets. For larger datasets, C2LSH is always the fastest technique since its having better algorithm time and number of disk seeks compared to the other two algorithms.  \\ 

\noindent\textbf{Query Processing Time (on SSD):} Figure \ref{fig:expTimeSSD} shows the overall time required to solve a given k-NN query on a Solid State Drive. In SSDs, I/O operations are much faster and the overall Query Processing Time is mainly dominated by the algorithm time. Therefore, C2LSH (which has the best Algorithm time) always performs the best on SSDs (for all datasets) followed by I-LSH (except for the smallest dataset, P53). \\

\noindent\textbf{Accuracy Ratio:} Figure \ref{fig:expRatio} shows the accuracy of the compared techniques. Having a ratio equal to 1 equates to highest accuracy. Except for the Mnist dataset, C2LSH produces the best accuracy among the three algorithms. QALSH is more accurate than I-LSH, which we believe is mainly because it uses more hash functions than I-LSH. Except for C2LSH's accuracy on the Mnist dataset, all three algorithms produce accurate results for all datasets. \\

\noindent Overall, we find that C2LSH can find k-NN results faster than QALSH and I-LSH. Additionally, all three algorithms produce accurate results (with C2LSH producing slightly better accurate results than QALSH and I-LSH for most datasets).

\section{Conclusion}
Approximate similarity search in high dimensional spaces has been an important problem in many diverse domains. In this paper, we focused on Locality Sensitive Hashing based techniques and presented a detailed experimental analysis on three famous LSH algorithms, C2LSH, QALSH, and I-LSH. For this analysis, we used various sizes of datasets and different yet important evaluation metrics. The results showed us that although a specific technique can perform better for smaller datasets but may not prove to be scalable and work well for larger datasets. We also observed that improvements in one portion of the LSH (e.g. I/O operations), do not results in overall improvements. Thus, trade-offs and different evaluation metrics should always be considered when comparing different techniques. In future we plan to also analyze the effect of changing the user-defined parameters on the performance of different techniques.

\end{document}